\documentclass[10pt, aps, prb, superscriptaddress, noshowkeys, twocolumn,
floatfix, preprintnumbers,showpacs]{revtex4-1}

\usepackage{amsmath, amssymb, amsfonts}
\usepackage{subfigure}
\usepackage{graphicx, color}
\usepackage{times}

\usepackage{graphicx}
\usepackage{bm,color}
\usepackage{times}
\usepackage{color}
\usepackage{ulem}

\usepackage[breaklinks]{hyperref}
\usepackage{bookmark} 
\hypersetup{colorlinks, citecolor = blue, filecolor = blue, linkcolor =
blue, urlcolor = blue}

\newcommand{\bk}{{\bf{k}}}

\newcommand{\bq}{{\bf{q}}}

\newcommand{\ud}{\text{d}}
\newcommand{\bA}{{\bf{A}}}

\renewcommand{\P}{{\bf{P}}}
\newcommand{\Ps}{\hat{\mathbf{\Psi}}}

\newcommand{\bnabla}{{\bf{\nabla}}}
\newcommand{\D}{{\bf{D}}}
\newcommand{\E}{{\bf{E}}}
\newcommand{\A}{{\bf{A}}}
\newcommand{\B}{{\bf{B}}}
\renewcommand{\H}{{\bf{H}}}
\newcommand{\epspar}{\epsilon_{\parallel}}
\newcommand{\Epar}{{\bf{E}_{\parallel}}}
\newcommand{\epsz}{\epsilon_{\bot}}
\allowdisplaybreaks

\begin{document}
\title{Influence of the effective layer thickness on the groundstate and excitonic properties of transition-metal dichalcogenide systems}

\author{L. Meckbach}
\affiliation{Department of Physics and Material Sciences Center,
Philipps University Marburg, Renthof 5, D-35032 Marburg, Germany}

\author{T. Stroucken}
\affiliation{Department of Physics and Material Sciences Center,
Philipps University Marburg, Renthof 5, D-35032 Marburg, Germany}

\author{S.W. Koch}
\affiliation{Department of Physics and Material Sciences Center,
Philipps University Marburg, Renthof 5, D-35032 Marburg, Germany}

\begin{abstract}

A self-consistent scheme for the calculations of the interacting groundstate and the near bandgap optical spectra of mono- and multilayer transition-metal-dichalcogenide systems is presented. The approach combines a dielectric model for the Coulomb interaction potential in a multilayer environment, gap equations for the renormalized groundstate, and the Dirac-Wannier-equation to determine the excitonic properties. To account for the extension of the individual monolayers perpendicular to their basic plane, an effective thickness parameter in the Coulomb interaction potential is introduced. Numerical evaluations for the example of MoS$_2$ show that the resulting finite size effects lead to significant modifications in the optical spectra, reproducing the experimentally observed non hydrogenic features of the excitonic resonance series. Applying the theory for multi-layer configurations, a consistent description of the near bandgap optical properties is obtained all the way from monolayer to bulk. In addition to the well-known in-plane excitons, also interlayer excitons occur in multilayer systems suggesting a reinterpretation of experimental results obtained for bulk material.

\end{abstract}
\date{\today}

\pacs{
}

\maketitle

\section{Introduction}

The optical and electronic properties of bulk transition-metal dichalcogenide systems (TMDCs) have been investigated intensively already in the 1970s\cite{beal1972,bordas1973,neville1976,fortin1975,anedda1979,anedda1980}. The excitonic series observed in the optical absorption spectra could be attributed to transitions at the $K$-points of the Brilliouin zone which nowadays are often referred to as Dirac points.\cite{beal1972,bordas1973,neville1976,fortin1975,anedda1979,anedda1980} However, as bulk TMDCs are indirect bandgap semiconductors, these materials have only played a minor role in the field of semiconductor optics in the following decades.
 
More recently, the interest in TMDCs and their optical properties has been revived with the ability to fabricate them as monolayers. Unlike their bulk counterparts, monolayers of several semiconducting TMDCs display a direct gap at the $K$-points of their respective Brillouin zone with a transition energy in the visible range\cite{kuc2011,yun2012,lambrecht2012,cappelluti2013,mak2010,zeng2013}. These systems exhibit a pronounced light-matter coupling and strong excitonic effects\cite{chernikov2014,he2014,ye2014,zhu2014}
The availability of different materials with a similar lattice structure but different bandgaps renders this material class extremely interesting as building blocks for heterostructures\cite{novoselov2016,dong2017}, and allows for the engineering of the overall electronic and optical properties to a wide extend.

For the systematic design and engineering of the electronic and optical properties of TMDC systems, it is highly desirable to have a predictive microscopic theory that includes the fundamental structural properties as well as the strong Coulomb interaction effects among the electronic excitations. In this article, we present a theoretical framework that allows us to determine both, the Coulombic renormalization of the $K$-point bandgap and the excitonic states. Our approach
combines a dielectric model to determine the Coulomb interaction potential in a multilayer environment, the gap equations for the renormalized ground state, and the Dirac-Wannier-equation  -- a generalization of the Mott-Wannier-equation  -- for the calculation of the excitonic states. 

Starting point of our theory is an effective two-band Hamiltonian, for which we use the massive Dirac-Fermion model (MDF)\cite{xiao2012}. Within the MDF, the gap equations and the Dirac-Wannier equation can be derived as static and linear part of the Dirac-Bloch equations,  i.e., the coupled equations of motion for the interband polarization and the electron-hole populations\cite{stroucken2017}. As our approach is based on the equations of motion approach, it can easily be extended to describe the nonlinear and dynamical optical properties.

In order to account for the finite out-of-plane monolayer extension, we introduce a thickness parameter $d$ in the effective Coulomb potential governing the interaction between the electronic excitations. The precise value of $d$ is determined by fitting a single spectral feature, e.g., the exact value of the energetically lowest excitonic resonance, to available experimental data. As all other parameters are extracted from first-principles density functional theory (DFT) calculations, $d$ is the only adjustable parameter in our theory. Once $d$ is fixed for a given material system, we are able to predict the bandgap and all the excitonic resonances for arbitrary dielectric environment and number of layers. Furthermore, we are able to study the optical properties for multi-layered structures and, in particular, the transition from a monolayer to bulk.

The paper is organized as follows: In Sec. \ref{sec:model}, we present the model system used for the calculations of the $K$-point groundstate and the optical properties of a multilayer structure.  In Sec. \ref{sec:Methods}, we derive the Wannier equation for the Dirac excitons and the gap equations that determine the renormalization groundstate properties. 
In Sec. \ref{sec:scaling}, we investigate finite size effects and the scaling properties of the coupled gap and Wannier equations for the simplified case of a constant background screening.  The results show that finite size effects lead to drastic modifications of the excitonic spectra.  
Finally, we analyze in Sec. \ref{sec:MLtobulk} the bandgap renormalization and near bandgap optical properties for mono- and multilayer configurations for the example of MoS$_2$, before we present a brief summary and discussion of our approach. In the appendix, we summarize important aspects of the electrostatic ingredients of our model, including the determination of the effective Coulomb interaction and screening properties.

\section{Model System}\label{sec:model}

Our model system is a stack of $N$ identical van-der-Waals bonded TMDC monolayers. Systematic studies of the bandstructure as function of the number of layers\cite{kuc2011,yun2012,lambrecht2012,cappelluti2013} show that the transition from direct to indirect occurs already when going from a monolayer to a bilayer configuration. This feature has been confirmed experimentally by layer-number dependent PL measurements\cite{mak2010,zeng2013} .  

At the same time, the DFT bandstructure investigations show that the bandstructure details around the $K$-points, which govern the optical absorption properties, are pretty much preserved while increasing the number of layers from monolayer to bulk\cite{kuc2011,yun2012,lambrecht2012,cappelluti2013,mak2010,zeng2013}. At the $K$ points, the out-of plane effective masses of the valence and conduction bands are typically much larger than those of the in-plane directions\cite{ye2015}. Consequently, the out-of-plane component of the kinetic energy can be neglected and the quasi-particles at the $K$-points can be considered as quasi-two dimensional particles well confined within the layers. Based on this observation,  we treat the $K$-point dynamics in a multilayer stack as $N$ electronically independent layers that are coupled via the Coulomb potential within the respective dielectric environment:
\[
H=\sum_n ( H_0^n+H_I^n) +\frac{1}{2}\sum_{nm,\bq}V_\bq^{nm}\hat\rho_\bq^n\hat\rho^m
_{-\bq}.
\]
Here $H_0^n$ describes the Hamiltonian of the $n^{th}$ layer, $H_I^n$ contains the light-matter interaction, and $H_C$ the Coulomb interaction, respectively. We assume that $\rho_\bq^n$, the charge density of the $n$-th layer, is strongly localized within that layer.
Treating the Hamiltonian of the isolated monolayer within an  effective two-band model, screening of the bands under consideration  is included dynamically, whereas the Coulomb matrix element $V_\bq^{nm}$ contains the screening of all the other bands and the dielectric environment.

\subsection{The Massive Dirac Fermion Hamiltonian}\label{sec:Hamiltonian}

According to {\it ab initio} methods based on DFT, the highest conduction and the lowest valence band are predominantly composed of $d$-type atomic orbitals of the metal atom\cite{mattheiss1973}.  Combining the relevant atomic orbitals that contribute to 
the valence and conduction bands into a two-component pseudo spinor, the minimal two-band Hamiltonian describing the near $K$-point properties in lowest order $\mathbf{k}\cdot\mathbf{p}$-theory can be written as\cite{xiao2012}
\begin{equation}
    \hat{H}_{0}^n = \sum_{s\tau,\mathbf{k}}\hat{\mathbf{\Psi}}^{\dagger}_{ns\tau\mathbf{k}} 
	  \left(at\mathbf{k}\cdot\hat{\sigma}_{\tau}+\frac{\Delta}{2}\hat{\sigma}_z 
	  -s\tau\lambda\frac{\hat{\sigma}_z-1}{2}\right)\hat{\mathbf{\Psi}}_{ns\tau\mathbf{k}}.
\end{equation}
Here, $\tau=\pm 1$ is the so called valley index, whereas $\Delta$, $2\lambda$, $t$ and $a$ denote the energy gap, the effective spin splitting of the valence bands, the effective hopping matrix element, and the lattice constant, respectively. The operator $\hat{\mathbf{\Psi}}_{ns\tau\mathbf{k}}$ is the tensor product of the electron spin state and the two component quasi-spinor in the $n$-th layer. The Pauli matrices $\hat{\sigma}_{\tau}=(\tau\hat{\sigma}_{x},\hat{\sigma}_{y})$
and $\hat{\sigma}_{z}$ act in the pseudo-spin space and $s$ is the $z$-component of the real spin, respectively. 
The eigenstates of $\hat{H}_{0}$ have the relativistic dispersion 
\begin{equation}
    \epsilon_{s\tau k} = \pm\frac{1}{2}\sqrt{\Delta_{s\tau}^2+(2\hbar v_F k)^2},\notag
\end{equation}
where $\Delta_{s\tau}=\Delta-s\tau\lambda$ denotes the spin and valley dependent energy gap at the K$^{\pm}$ points and $v_F=at/\hbar$ is the Fermi-velocity.

Employing the minimal substitution principle, the light-matter (LM) Hamiltionian is obtained as 
\begin{equation}
H_I^n =-e\frac{v_F}{c}\sum_{s\tau\bk}\Ps^\dagger_{ns\tau\bk}{\bA}^n \cdot {\hat{\bf
\sigma}}_\tau\Ps_{ns\tau\bk}.
\end{equation}
Expanding the charge density in terms of the pseudo spinors, we find for the Coulomb interaction
\[
H_C=\frac{1}{2}\sum_{nm\bk\bk'\bq}:\Ps^\dagger_{ns\tau\bk-\bq}\Ps_{ns\tau\bk}V_\bq^{nm}\Ps^\dagger_{ms\tau\bk'+\bq}\Ps_{ms\tau\bk'}:
\]
where $:\cdot :$ denotes normal ordering.

\subsection{Coulomb Potential in a Multilayer Environment}\label{sec:Coulomb}

\begin{figure}[hbt]
  \includegraphics[width = 0.33\textwidth]{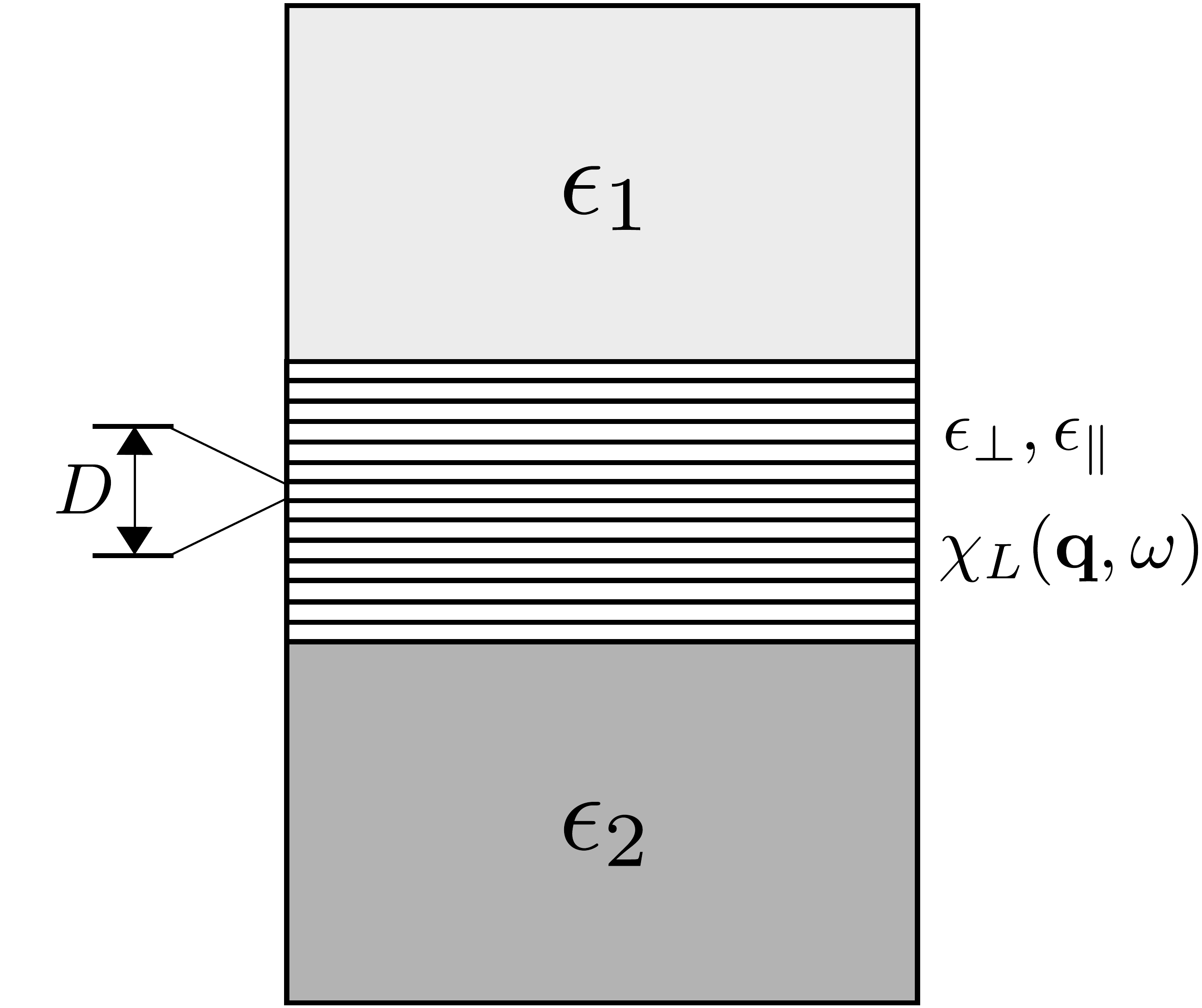}
  \caption{\label{fig:schematics} Schematic of the model system. The distance between the van der Waals bonded layers is denoted by $D$.}
\end{figure}

The Coulomb interaction potential in our two-band Hamiltonian
contains screening contributions from the system's environment, such as substrate screening etc., and possible non-resonant intrinsic contributions arising from all other bands. To avoid double counting, it is important to separate the contributions of the explicitly treated bands from the rest. Since the DFT dielectric tensor contains all the ingredients, the separation of resonant and non-resonant contributions is a nontrivial task.

Here, we develop a scheme that combines bulk DFT calculations of the dielectric tensor with analytical results obtained within the MDF model that allows us to determine the fully screened and non-resonantly screened ('bare') Coulomb potential for various dielectric environments.
To derive the Coulomb interaction potential in the multilayer environment, we start from Maxwell's equations
\begin{eqnarray}
\bnabla\cdot\D&=&4\pi \rho_{ext}\\
\bnabla\cdot\B&=&0\\
\bnabla\times\H-\frac{1}{c}\dot\D&=&\frac{4\pi }{c}{\bf j}_{ext}\\
\bnabla\times\E+\frac{1}{c}\dot\B&=&0.
\end{eqnarray}
For the layered material, we make the ansatz
\begin{equation}
\B=\H,\quad\quad \D=\epspar\Epar+\epsz E_z{\bf e}_z+4\pi\P,
\end{equation}
where $\epspar\equiv \epspar(z)$ and $\epsz\equiv\epsz(z)$ represent the  non-resonant contributions to the anisotropic 
dielectric tensor and $\P$ contains all nonlocal, time and frequency dependent resonant contributions. 
The non-resonant contributions are assumed to be local in space and time and constant within a slab of thickness $L=ND$, where $N$ is the number of layers and $D$ the natural layer-to-layer distance in the bulk parent material (see Fig. \ref{fig:schematics}). 

As the considered structure is homogeneous with respect to the in-plane coordinates but inhomogeneous with respect to the out-of-plane coordinates, we use a mixed $(\bq,z)$ representation in the following, where $\bq$ is the in-plane wave vector. With $\B=\bnabla\times\A$, $\E=-\dot\A/c-\bnabla\phi$ and  the  generalized Coulomb gauge
$
\epspar\bnabla_\parallel\cdot\A_{\parallel}+\epsz\partial_z A_z=0
$, a division into in-plane transverse and longitudinal contributions yields Poisson's equation for the scalar potential
\begin{eqnarray}
\label{2DCoulomb}
\left(-\epsz\partial^2_z+\epspar\bq^2\right)\phi
=4\pi\left(\rho_{ext}- i \bq\cdot \P_\parallel^{L}-\partial_zP_z\right).
\label{Eq:Poisson}
\end{eqnarray}
The solution of this equation for a $\delta$-inhomogeneity $\rho_{ext}=\delta(z-z')$ and $\P_\parallel^L=0$ determines the 'bare' Coulomb potential $V_\bq(z,z')$. Correspondingly, the screened Coulomb potential is obtained as solution of Poisson's equation with resonant contributions.
 Provided the non-resonant contributions to the dielectric tensor are known,  the bare Coulomb interaction can be obtained analytically
from Eq. (\ref{Eq:Poisson}).

For the resonant contributions to the longitudinal polarization, we assume that these are composed of 
 a sum of  localized (2D)  parts, that are treated within linear response. In the strict 2D limit, these can be expressed as
\begin{equation}
\P=-ie^2\bq\sum_{n=1}^N \chi_L(\bq,\omega)\phi(\bq,z_n,\omega)
\delta(z-z_n),
\label{Eq:PolAnsatz}
\end{equation}
where $z_n=(n-1/2)D$ is the central position of the $n^{th}$ layer and  $\chi_L(\bq,\omega)$ is the longitudinal susceptibility, respectively. The longitudinal susceptibility is related to the polarzation function of the 2D layer via $\chi_L(\bq,\omega)=-\Pi(\bq,\omega)/q^2$.
 Within the MDF model, for each spin and valley combination, the long-wavelength limit of the static RPA polarization function gives \cite{rodin2013}
\[
\Pi(\bq,0)=-\frac{1}{6\pi}\frac{q^2}{\Delta_{s\tau}}
\]
where $\Delta_{s\tau}$ is the spin and valley dependent gap at the Dirac points. Summing over the spin and valley indices, one finds
\[
r_0=\lim_{\bq\rightarrow 0}2\pi e^2\chi_L(\bq,0)=\frac{2 e^2(\Delta_A+\Delta_B)}{3\Delta_A\Delta_B},
\]
which is of the order of $10$ \AA\  for a typical MX$_2$ monolayer, independent of the dielectric environment.

Inserting Eq. (\ref{Eq:PolAnsatz}) into Eq. (\ref{Eq:Poisson}), we obtain for the screened Coulomb interaction
\begin{eqnarray}
 V_{S,\bq}^{nm}(\omega)=\sum_{l=1}^N\left(\delta_{nl}+e^2q^2\chi(\bq,\omega)V_{\bq}^{nl}\right)
 ^{-1}V_\bq^{lm}.
\label{VCmatrix}
\end{eqnarray}
Eq.  (\ref{VCmatrix}) expresses the screened Coulomb interaction in terms of the bare potential and an inverse nonlocal dielectric function. 
With the aid of the screened and unscreened interaction, we can define the local dielectric functions 
$\epsilon^n(\bq,\omega)=V_{{\rm Vac},\bq}^{nn}/V_{S,\bq}^{nn}(\omega)$, where $V_{{\rm Vac},\bq}^{nn}=2\pi /|\bq|$ is the 2D Coulomb potential in vacuum. Similarly, we introduce the resonant and nonresonant contributions of the local dielectric functions as 
$\epsilon^n_{\rm res}(\bq,\omega)=V_{\bq}^{nn}/V_{S,\bq}^{nn}(\omega)$ and $\epsilon^n_{\rm nr}(\bq,\omega)=V_{{\rm Vac},\bq}^{nn}/V_{\bq}^{nn}$, respectively. In general, each layer within the multilayer environment has a different local dielectric function reflecting its respective dielectric environment.

For a bulk 
material consisting of $N\gg 1$ regularly spaced layers, Eq. (\ref{2DCoulomb}) can be solved by a  Fourier transformation, giving
\[
V_S(\bq,q_z)=\frac{4\pi }{\epsz q_z^2+
\bq^2\left(\epspar+4\pi e^2\chi_L(\bq,\omega)/D\right)},\]
where $D$ is the layer-to-layer distance.
Comparison with the 3D anistropic Coulomb interaction suggests that the bulk in-plane dielectric 
constant is given by
\[
\epspar^B=\epspar+\lim_{\bq\rightarrow 0}4\pi e^2\chi_L(\bq,\omega)/D.
\]
 We use this relation and the bulk values for the macroscopic background dielectric constants obtained by DFT \cite{ghosh2013} to determine the required values of  $\epspar$ and $\epsz$. 

\subsection{Quasi-2D Coulomb Potential}

Computing the Coulomb potential for a strictly 2D layer ignores the fact that the spatial carrier distribution in the out-of-plane direction has a finite extension and is not a sharp $\delta$-function at the central layer position. Hence, instead of solving Poisson's equation with a $\delta$-singulartity, we have to compute the scalar potential for a charge distribution $\rho_\bq(z-z_n)$ induced by the charge density in the $n^{th}$ layer and replace Eq. (\ref{Eq:PolAnsatz}) by (see Appendix \ref{App:Poisson})
\begin{eqnarray}
\P&=&-ie^2\bq\sum_{n=1}^N \chi_L(\bq,\omega)\rho_\bq(z-z_n)
\nonumber\\
&\times&\int_{-D/2}^{D/2} dz'\phi(\bq,z',\omega)\rho_{-\bq}(z'-z_n).
\end{eqnarray}
Defining the quasi-2D Coulomb potential between different layers as
\[
\bar V_\bq^{nm}=\int_{-D/2}^{D/2}dz\int_{-D/2}^{D/2}dz'\rho_{-\bq}(z'-z_n)V_\bq(z,z')\rho_{\bq}(z-z_m)
\]
and similar for the screened interaction potential, Eq. (\ref{VCmatrix}) remains valid with all matrix elements replaced by the quasi-2D ones. 

In order to have a simple expression, we use in in the following the 2D Ohno potential 
\[
 \bar V_\bq^{nm}\approx V_\bq^{nm}{\rm e}^{-qd},
\]
as approximation for the bare quasi-2D potential. Here, $d$ denotes the effective thickness parameter accounting for finite out-of-plane size effects.

\section{Methods}\label{sec:Methods}

The Coulomb interaction leads to renormalizations of the single-particle bandstructure and to excitonic effects in the optical properties of a semiconductor. In this section, we follow the derivation in Ref. \onlinecite{stroucken2017} to  show how  both of these features are obtained within the equations of motion (EOM) approach. Here, one derives the equations of motion for the interband polarization and the valence and conduction band occupation probalities to obtain
the semiconductor Bloch equations (SBE)\cite{haugkoch2009} which describe excitonic effects as well as the excitation dependent energy renormalizations. 

As input for the SBE, one needs the single-particle bandstructure and the system's groundstate properties. Since DFT-based bandstructure calculations usually underestimate the unexcited bandgap, one often uses the experimental values instead of the DFT results. Whereas this approach works well for the typical GaAs-type bulk or mesoscopic semiconductor structures, the fundamental gap of mono- or few-layer TMDCs is experimentally difficult to access and depends strongly on the dielectric environment. Therefore, it is desirable to compute the gap renormalization self-consistently from first principles.

\subsection{Gap Equations}

As shown in Ref. \onlinecite{stroucken2017}, the combination of the EOM with a variational approach yields a set of coupled integral equations --the gap equations-- for the renormalized bandgap and the Fermi velocity. The gap equations are non-perturbative and can be derived on the same level of approximation as the EOM for the excitation dynamics. We define the dynamical variables
\begin{eqnarray}
\Gamma_{s\tau\mathbf{k}} &=& f^{b}_{s\tau\mathbf{k}} - f^{a}_{s\tau\mathbf{k}} 
	= \langle\hat{b}^{\dagger}_{s\tau\mathbf{k}}\hat{b}_{s\tau\mathbf{k}}\rangle
	- \langle\hat{a}^{\dagger}_{s\tau\mathbf{k}}\hat{a}_{s\tau\mathbf{k}}\rangle,\\
\Pi_{s\tau\mathbf{k}} &=& \langle\hat{b}^{\dagger}_{s\tau\mathbf{k}}\hat{a}_{s\tau\mathbf{k}}\rangle,
\end{eqnarray}
where $\hat{a}^\dagger_{s\tau\mathbf{k}}$ and $\hat{b}^\dagger_{s\tau\mathbf{k}}$ create a particle in the basis states spanning the pseudo-spinor $\Ps^\dagger_{s\tau\bk}$.
Since the groundstate should be static, we search for the stationary solutions of Heisenberg's equations of motion,
\begin{eqnarray}
i\hbar\frac{d}{dt}\Pi_{s\tau\mathbf{k}}  &=& \left(\Delta_{s\tau} + \hat{V}[\Gamma_{s\tau}]\right)\Pi_{s\tau\mathbf{k}}\notag\\ 
					 &+& \left(\tau\hbar v_{F} k e^{-i\tau\theta_{\mathbf{k}}}-\hat{V}[\Pi_{s\tau}]\right)\Gamma_{s\tau\mathbf{k}},\\
i\hbar\frac{d}{dt}\Gamma_{s\tau\mathbf{k}} &=& 2\Pi_{s\tau\mathbf{k}}\left(\tau\hbar v_{F} k e^{i\tau\theta_{\mathbf{k}}}-\hat{V}[\Pi^{*}_{s\tau}]\right)\notag\\
					   &-& 2\Pi^{*}_{s\tau\mathbf{k}}\left(\tau\hbar v_{F} k e^{-i\tau\theta_{\mathbf{k}}}-\hat{V}[\Pi_{s\tau}]\right)
\end{eqnarray}
in the absence of an externally applied optical field. To simplify the notatation, we introduced the functional relation $\hat{V}[f]\equiv\sum_{\mathbf{k'}}\,V_{\mathbf{|k-k'|}}\,f_{\mathbf{k'}}$. Demanding a stationary solution, we find 
\begin{eqnarray}
\label{Eq:Pistat}
0 &=& \tilde{\Delta}_{s\tau\mathbf{k}}\Pi_{s\tau\mathbf{k}} + \tau\hbar \tilde{v}_{s\tau\mathbf{k}} k 
	e^{-i\tau\theta_{\mathbf{k}}}\Gamma_{s\tau\mathbf{k}},\\
0 &=& \Im\left[\Pi_{s\tau\mathbf{k}}\tau\hbar \tilde{v}_{s\tau\mathbf{k}} k 
	e^{i\tau\theta_{\mathbf{k}}}\right],
\label{Eq:Gammastat}
\end{eqnarray}
where
\begin{eqnarray}
\label{Eq:Delta}
\tilde{\Delta}_{s\tau\mathbf{k}} &=& \Delta_{s\tau} + \hat{V}[\Gamma_{s\tau}],\\
\tau\hbar\tilde{v}_{s\tau\mathbf{k}}k e^{-i\tau\theta_{\mathbf{k}}} &=& \tau\hbar v_{F} k e^{-i\tau\theta_{\mathbf{k}}} - \hat{V}[\Pi_{s\tau}] 
\label{Eq:vf}
\end{eqnarray}
are the renormalized bandgap energy and Fermi-velocity, respectively.
Together with the relation $1=\Gamma^{2}_{s\tau\mathbf{k}}+4 |\Pi_{s\tau\mathbf{k}}|^2$, which holds for any coherent state, 
we obtain the  algebraic equations
\begin{eqnarray}
\label{Eq:Pi}
\Pi_{s\tau\mathbf{k}} &=& -\frac{\tau\hbar\tilde{v}_{s\tau\mathbf{k}}k}{2\tilde{\epsilon}_{s\tau\mathbf{k}}}e^{-i\tau\theta_{\mathbf{k}}},\\
\Gamma_{s\tau\mathbf{k}} &=&  \frac{\tilde{\Delta}_{s\tau\mathbf{k}}}{2\tilde{\epsilon}_{s\tau\mathbf{k}}}
\label{Eq:Gamma}
\end{eqnarray}
with
\begin{equation}
\tilde{\epsilon}_{s\tau\mathbf{k}}=\frac{1}{2}\sqrt{\tilde{\Delta}_{s\tau\mathbf{k}}^2+\left(2\hbar\tilde{v}_{s\tau\mathbf{k}} k\right)^2}.
\end{equation}
Inserting Eqs. (\ref{Eq:Pi}) and (\ref{Eq:Gamma}) into Eqs. (\ref{Eq:Pistat}) and (\ref{Eq:Gammastat}) yields the closed set of integral equations, the gap equations, as
\begin{eqnarray}
\tilde{\Delta}_{s\tau\mathbf{k}} &=& \Delta_{s\tau}+\frac{1}{2}\sum_{\mathbf{k'}}\,V_{\mathbf{|k-k'|}}\,		
	\frac{\tilde{\Delta}_{s\tau\mathbf{k'}}}{\tilde{\epsilon}_{s\tau\mathbf{k'}}},\nonumber\\
\tilde{v}_{s\tau\mathbf{k}} &=& v_F + \frac{1}{2}\sum_{\mathbf{k'}}\,V_{\mathbf{|k-k'|}}\,
	\frac{k'}{k}\frac{\tilde{v}_{s\tau\mathbf{k'}}}{\tilde{\epsilon}_{s\tau\mathbf{k'}}} 
	e^{i\tau(\theta_{\mathbf{k}}-\theta_{\mathbf{k'}})}.
	\label{Eq:Gap}
\end{eqnarray}

It is easily verified that 
$\tilde{\Delta}_{s\tau\mathbf{k}}$ and $\tilde{v}_{s\tau\mathbf{k}}$ define the mean-field Hamiltonian
\begin{equation}
\hat{H}^{MF} = \sum_{s,\tau,\mathbf{k}}\hat{\mathbf{\Psi}}^{\dagger}_{s\tau \mathbf{k}}
	\left(\hbar\tilde{v}_{s\tau\mathbf{k}}\mathbf{k}\cdot\hat{\boldsymbol{\sigma}}_{\tau}
	+\frac{\tilde{\Delta}_{s\tau\mathbf{k}}}{2}\hat{\sigma}_z\right)
	\hat{\mathbf{\Psi}}_{s\tau \mathbf{k}}
\end{equation}
with the eigenvalues $\pm \tilde{\epsilon}_{s\tau\mathbf{k}}$.
The corresponding eigenstates are given by
\begin{eqnarray}
\Psi_{\bk}^c=\left(\begin{array}{c}u_{s\tau k}\\v_{s\tau k}{\rm e}^{i\tau\theta_\bk}\end{array}\right),\quad\Psi_{\bk}^\nu=\left(\begin{array}{c}v_{s\tau k}{\rm e}^{-i\tau\theta_\bk}\\-u_{s\tau k}\end{array}\right),
\label{Eq:spinors}
\end{eqnarray}
where
 $u_{\tau k}=\sqrt{(\tilde\epsilon_{s\tau k}+\tilde\Delta_{s\tau k}/2)/2\tilde\epsilon_{s\tau k}}$ and  $v_{s\tau k}=\sqrt{(\tilde\epsilon_{s\tau k}-\tilde\Delta_{s\tau k}/2)/2\tilde\epsilon_{s\tau k}}$.
As usual in intrinsic semiconductors, the groundstate is characterized by a completely filled valence and empty conduction band, respectively. Since ${\varepsilon}_{s\tau\bk}>\epsilon_{s\tau\bk}$, the total energy lies below the energy of the non-interacting groundstate.

\subsection{Dirac-Bloch and Dirac-Wannier Equations}\label{secDBE}

To determine the excitation dynamics of our model system, we transform the Hamiltonian into the electron-hole picture using the {\it renormalized} bandstructure
and eigenstates. Furthermore, we use the interband transition amplitudes and occupation numbers of the renormalized bands as dynamical variables,
\begin{eqnarray}
P_{s\tau\bk}&=&\langle\nu^\dagger_{s\tau\bk}c_{s\tau\bk}\rangle,\\
f_{s\tau\bk}&=&1- \langle \nu^\dagger_{s\tau\bk} \nu_{s\tau\bk}\rangle =\langle c^\dagger_{s\tau\bk} c_{s\tau\bk} \rangle.
\end{eqnarray}
It is easily verified that, using the renormalized bands, the groundstate expection values are given by $P_{s\tau\bk}=f_{s\tau\bk}=0$ (note: this is not true for the transition amplitudes and occupation numbers within the unrenormalized bands!).

At the Hartree-Fock level, the resulting Heisenberg EOM for the dynamical variables are given by\cite{stroucken2017}:
\begin{eqnarray} \label{GBEpol}
i\hbar \frac{\ud}{\ud t} P_{s\tau\bk}
&=& 2 \left(\Sigma_{s\tau\bk}-\frac{1}{c}\bA\cdot\bf{j}_{s\tau\bk}\right) P_{s\tau\bk}\nonumber\\ 
&-& (1-2f_{s\tau\bk})\Omega_{s\tau\bk}- \left. i\hbar\frac{\ud}{\ud t} P_{s\tau\bk} \right|_{\text{coll}},\\
 \label{GBEpop}
\hbar \frac{\ud}{\ud t} f_{s\tau\bk} 
&=& -2\Im \left[ P^*_{s\tau\bk} \Omega_{s\tau\bk} \right]
-\left. \hbar\frac{\ud}{\ud t} f_{s\tau\bk} \right|_{\text{coll}} .
\end{eqnarray}
In these {\it Dirac-Bloch equations} (DBE), the Coulomb interaction leads to excitation dependent renormalizations of the single-particle energy and the Rabi frequency, 
\begin{widetext}
\begin{eqnarray}
 \Sigma_{s\tau\bk}
&=&\tilde \epsilon_{s\tau\bk}-\sum_{\bk'}V_{|\bk-\bk'|} \bigl[ W_{cccc}(\bk,\bk')-W_{c\nu\nu c}(\bk,\bk') \Bigr] f_{s\tau\bk'}
+\sum_{\bk'} V_{|\bk-\bk'|}\bigl[W_{cc\nu c}(\bk,\bk')P_{s\tau\bk'} +c.c. \Bigr],
 \label{erenorm}\\
\Omega_{s\tau\bk}
&=& \sum_{\bk'} V_{|\bk-\bk'|}
\left[W_{cc\nu\nu}(\bk,\bk')P_{s\tau\bk'} + W_{c\nu c\nu}(\bk,\bk') P^*_{s\tau\bk'}
 -2 W_{c\nu\nu\nu}(\bk,\bk')f_{s\tau\bk'}\right]
\nonumber\\&+&
\tau\sqrt{2}\frac{e v_F}{c}\left(v_k^2{\rm e}^{-2i\tau\theta_\bk}A^\tau-u_k^2A^{-\tau}\right).
\label{rabi}
\end{eqnarray}
\end{widetext}
whereas groundstate renormalizations are contained in the renormalized dispersion $\tilde \epsilon_{s\tau\bk}$.
Here,
\[
W_{\alpha\alpha'\beta\beta'}(\bk,\bk')=\langle \alpha\bk|\alpha'\bk'\rangle\langle \beta\bk'|\beta'\bk\rangle
\] 
contains the overlap matrix elements between the renormalized conduction and valence bands.
Despite the formal equivalence of Eqs. (\ref{GBEpol}) and (\ref{GBEpop}) to the standard SBE, the renormalized single-particle energy and Rabi frequency differ from the standard expressions  by the Coulomb matrix elements for scattering processes across the bands, i.e. Auger-type processes and electron-hole pair creation and annihilation.
In Eqs. (\ref{GBEpol}) and (\ref{GBEpop}), the terms $\ud / \ud t |_\text{coll}$ refer to incoherent scattering contributions 
beyond the Hartree-Fock approximation and ${\bf j}_{s\tau\bk}=-\tau\frac{e}{\hbar}\nabla_{\bk}\tilde\epsilon_{s\tau\bk}$
is the intraband current matrix element, respectively.

The Dirac-Wannier equation (DWE) is obtained from the DBE as homogeneous part of the linearized polarization equation,
\begin{widetext}
\begin{eqnarray}
2 \tilde\varepsilon_{s\tau\bk}\phi_{s\tau\lambda}(\bk) 
& -&\sum_{\bk'} V_{|\bk-\bk'|}\left[W_{cc\nu\nu}(\bk,\bk')\phi_{s\tau\lambda}(\bk') + W_{c\nu c\nu}(\bk,\bq) \phi^*_{s\tau\lambda}(\bk')\right]
=E_{s\tau\lambda}\phi_{s\tau\lambda}(\bk) \, .
\label{WannierDirac}
\end{eqnarray}
\end{widetext}
Apart from the dispersion, the DWE differs from the standard Mott-Wannier
equation by the last term on the l.h.s. of Eq. (\ref{WannierDirac}), that describes a coupling of the $\phi$ and $\phi^*$ by spontaneous pair creation and annihilation.
In view of the large gap in semiconducting TMDCs, these contributions are frequently neglected. However, the validity of this approximaton is not a priori clear since is actually depends on the strength of the Coulomb interaction. In our evaluations in this paper, we therefore avoid the wide-gap approximation (WGA).

\section{Finite Thickness Effects}\label{sec:scaling}

In the strict 2D limit, the exciton binding and wavefunctions at the origin become singular in the regime of strong Coulomb interactions\cite{rodin2013,stroucken2015} leading to an excitonic collapse of the interacting groundstate. In this case, the system undergoes a transition into an excitonic insulator state, where the bright optical resonances correspond to intra-excitonic transitions of a BCS-like excitonic condensate\cite{stroucken2015,stroucken2017}. A similar divergence of the binding energy and wavefunctions is known in QED for hydrogen-like atoms with $Z>137$. In QED, this "catastrophe" is treated via a regularization of the Coulomb-potential accounting for a small but finite extension of the nucleus, i.e., by replacing the $1/r$ potential by the Ohno potential $1/\sqrt{r^2+d^2}$. 

In this section, we apply a similar procedure and investigate the influence of finite size effects on the gap and exciton equations for a monolayer with a constant background screening $\kappa$, i.e. $\bar V_\bq=2\pi  e^2 {\rm e}^{-qd}/\kappa q$. This potential is appropriate for both, a monolayer embedded in bulk with $\kappa=\sqrt{\epspar\epsz}$ and for the long wavelength limit $qD\rightarrow 0$ of a monolayer on a substrate with $\kappa=(\epsilon_S+1)/2$ (see Appendix).

In order to unify the description of different material systems and to identify the general aspects of the obtained results, it is often advantageous to introduce scaled units. For the problem under investigation here, one can either choose relativistic or 
excitonic units. As the only absolute energy value entering into the DWE,
one can use the single-particle gap $\Delta$ as energy unit. The single-particle dispersion is then found as $\epsilon_k/\Delta=\pm\frac{1}{2}\sqrt{1+(k\lambda_C)^2}$, 
where $\lambda_C=2\hbar v_F/\Delta$ is the Compton wavelength of the electrons and holes. Using the Compton wavelength as length scale, the scaled quasi-2D Coulomb potential
is given by
\begin{equation}\label{eq:CoulombInteraction}
    \overline{\bar V}_{\overline{\mathbf{q}}} = \frac{\bar V_{\mathbf{q}}}{\Delta}
	  = \frac{\pi\alpha}{\bar q} {\rm e}^{\overline{q}\overline{d}},
\end{equation}
which is characterized by the  parameter combination $\alpha=e^2/\kappa\hbar v_F$.  The Compton wavelength allows one to distinguish between the relativistic and the non-relativistic regimes, where 
the latter one is found on a length scale large compared to the Compton wavelength. 

Using scaled units, it is easily shown that the total Hamiltonian is characterized by two parameters, namely the effective fine structure constant $\alpha$ and the effective thickness parameter $d$. Consequently,  both the gap equations and the exciton equation are
characterized by the same parameters. The long-wavelength limit of the resonant part of the RPA dielectric function in scaled units is obtained as
 \begin{equation}
\epsilon_{\rm res}(\bq)=1+\frac{2}{3}\alpha q\bar\lambda_C{\rm e}^{-qd},\notag
\end{equation}
where $\bar\lambda_C=(\lambda^A_C+\lambda^B_C)/2$ is the avarage of the respective Compton wavelengths associated with the gap of the $A$ and $B$ excitons. This dielectric function 
is of a similar form as the  potential first introduced by Keldysh\cite{keldysh1979} for a thin sheet with constant sheet polarizability and has been used by several authors\cite{cudazzo2011,pulci2012,berkelbach2013,chernikov2014,wu2015,latini2015} to model the 
excitonic properties of TMDCs. As a consequence of the Ohno potential, the dielectric function does not increase to infinity with increasing $q$ but approaches its 
maximum value at $q=1/d$. A similar behavior has been found by first principle calculations including finite size effects\cite{latini2015} or using a truncated Coulomb potential\cite{qiu2016}. Furthermore, the screening length $\frac{2}{3}\alpha \bar\lambda_C$ contains resonant contributions only.

When discussing excitonic properties, it is sometimes useful to resort to excitonic units.
The Compton wavelength and the (3D) exciton Bohr radius  $a_0=\hbar^2\kappa/m_r e^2$ are related via $a_B=2\lambda_C/\alpha$, and the exciton Rydberg $Ry=m_re^4/2\hbar^2\kappa^2$ 
is related to the gap via $Ry=\alpha^2\Delta/8$, respectively. In the following, we will use both unit systems in order to emphasize systematic dependencies and 
the essential underlying physics.    

\subsection{Numerical Solution of the Gap Equations}

\begin{figure}[hbt]
  \includegraphics[width = \columnwidth]{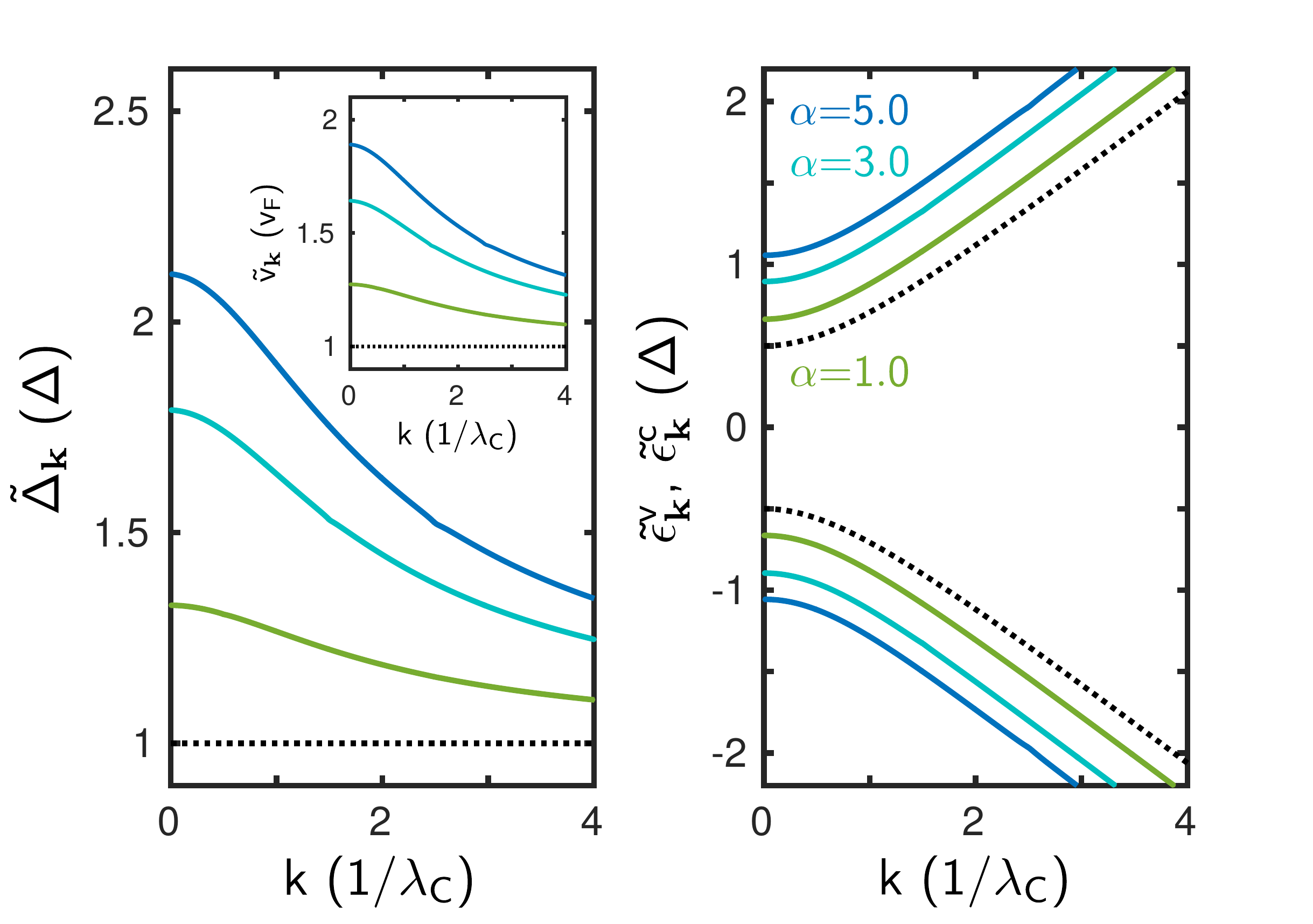}
  \caption{\label{fig:WIGS_GE} Renormalized band gap energy (left) and Fermi-velocity (inset) distributions for $d=1.0\lambda_C$ and $\alpha=1.0$ (green), 3.0 (cyan), and 5.0 (blue). The resulting renormalized single-particle dispersion is shown on the right. The black dotted lines represent the non-interacting ground state properties.}
\end{figure}

Examples of our numerical solutions of the gap equations (\ref{Eq:Gap}) are shown in Fig. \ref{fig:WIGS_GE}. Here, we plot $\tilde{\Delta}_{\mathbf{k}}$ and $\tilde{v}_{\mathbf{k}}$ as well as the resulting renormalized single-particle 
dispersion $\tilde{\varepsilon}_{\mathbf{k}}$ for various values of $\alpha$ and a fixed thickness parameter $d=1.0\lambda_C$. Both $\tilde{\Delta}_{\mathbf{k}}$ (left) and $\tilde{v}_{\mathbf{k}}$ (inset) have their maxima at $k=0$  and  converge to their respective non-interacting groundstate values $\Delta$ and $v_F$ (respective black dotted lines) for large $k$.
Within a good approximation, the 
renormalization of the band gap energy and the Fermi velocity leads to a rigid  shift of the non-interacting single-particle dispersion (right panel in Fig. \ref{fig:WIGS_GE}), in agreement with reported predictions based on the \textsl{GW} approximation\cite{komsa2012,shi2013,rasmussen2015,qiu2016}.

\begin{figure}[hbt]
  \includegraphics[width = \columnwidth]{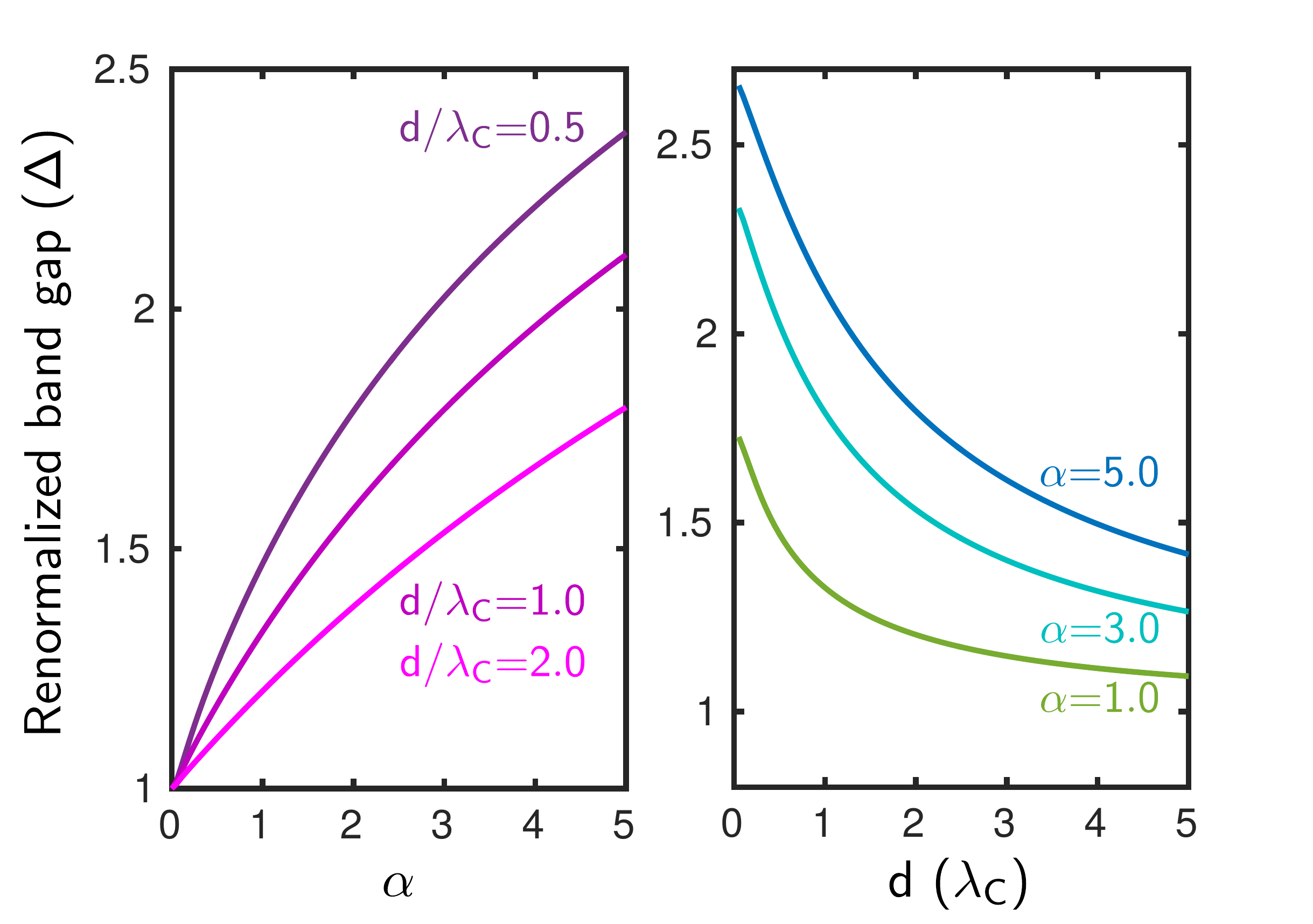}
  \caption{\label{fig:WIGS_GE_2}Left:  Renormalized gap as function of coupling constant $\alpha$ for three different thickness parameters $d=.5\lambda_C$ , $d=1.0\lambda_C$ and $d=2\lambda_C$. Right: Dependence of the renormalized gap on the effective thickness parameter $d$ for two values of the coupling constant $\alpha$.}
\end{figure}

Since the renormalization does not lead to a deformation of the single-particle bandstructure, it only shifts the energetic position of the excitonic resonances in the respective optical spectra but does not influence their binding energies. Hence, it suffices to study the overall gap shift as function of the system parameters $\alpha$ and $d$. For this purpose, we plot in the left panel of Fig. \ref{fig:WIGS_GE_2} the computed dependence of the renormalized gap on $\alpha$ for three different values of the effective thickness parameter. As we can see, the gap increases linearly with $\alpha$ for small values of the coupling strength switching over to a logarithmic increase for large coupling strengths.

In the right panel of Fig. \ref{fig:WIGS_GE_2}, we show the computed values of the renormalized gap as function of the effective thickness parameter $d$ for three different values of $\alpha$. 
We notice a sensitive $d$ dependence of the gap in the region where $d\lesssim  \alpha\lambda_C$, which is typically realized in TMDC structures.

\subsection{Numerical Solution of the Dirac-Wannier Equation}

Often \cite{berkelbach2013,chernikov2014,latini2015}, the excitonic properties of 
TMDCs are treated in the WGA where the relativistic quasi-particle dispersion can be approximated by parabolic bands and all 
contributions $\propto v_kv_{k'}$ in the Coulomb matrix elements can be neglected. As a result, the excitonic states become independent of the Compton wavelength and the only 
remaining length scales are the effective sheet thickness $d$ and the exciton Bohr radius $a_B$. Moreover, states with $m=\pm|m|$ are degenerate. 

Since the only energy scale 
other than the gap is the exciton Rydberg energy, the WGA is actually equivalent to the nonrelativistic approximation $\alpha\ll 1$. For typical TMDC parameters, the effective coupling 
constant is in the range of $\alpha\propto 3/\kappa-5/\kappa$, clearly questioning the WGA. 
Corrections to the WGA result both from the full relativistic dispersion and from the lifting of the degeneracy between states with opposite orbital angular momentum\cite{wu2015,zhou2015}

\begin{figure}[hbt]
  \includegraphics[width = \columnwidth]{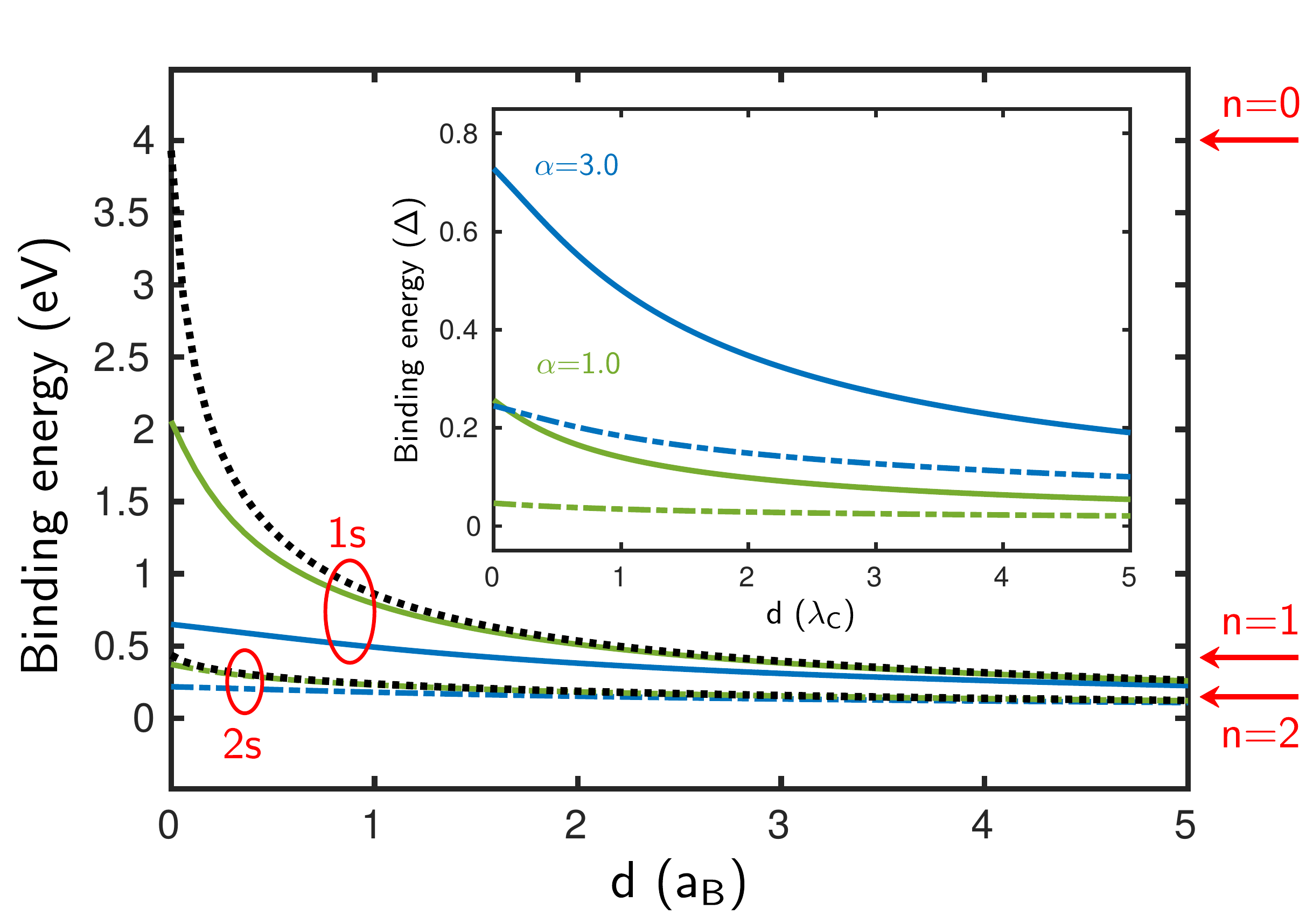}
  \caption{\label{fig:E_1s_2s_ExcUnits} Binding energy of the 1s-  and 2s-exciton in dependence of the effective thickness parameter $d$
     for $\alpha=1.0$ (green) and $\alpha=3.0$ (blue). 
     The black dotted lines show the non-relativistic case for parabolic bands. The strict 2D non-relativistic limit ($d=0$) is marked by the arrows.}
\end{figure}

Numerically solving the full DWE (\ref{WannierDirac}) we obtain the results shown in    
Fig. \ref{fig:E_1s_2s_ExcUnits}. Here, we plot the binding energies of the 1s-(solid lines) 
and 2s-exciton (dashed lines) as functions of the effective thickness parameter in excitonic units for $\alpha = 1.0$ and $\alpha = 3.0$. For reference, the arrows mark the binding of the  exciton states with main quantum number $n=0$, $n=1$ , and $n=2$  within the 2D hydrogen model. 

For finite values for the effective sheet thickness $d$, the Coulomb interaction close to the origin is weakened relative to the strict 2D case, affecting particularly the strongest bound $s-$type excitons with large probability density at the origin. Fig. \ref{fig:E_1s_2s_ExcUnits} clearly shows that the binding energies of the $1s$ and $2s$ excitons vary strongly with the sheet thickness in the regime where $d \approx a_B$ and become pretty much $d$ independent for $d \gg a_B$. 
In that limit, the binding energy of the  $1s-$ exciton drops below the value of the $n=1$ 2D-exciton state. At the same time, the $2s$ binding energy seems to converge toward the $n=2$ value of the 2D limit  leading  to an overall strongly non-hydrogenic behavior of the exciton series similar to the experimental observations\cite{chernikov2014,he2014,ugeda2014,wang2015}.
This behavior is quite different from what is known for semiconductor quantum wells, where the exciton series changes from a 2D to 3D Rydberg series if the sample dimensions exceed the exciton Bohr radius.

\begin{figure}[hbt]
  \includegraphics[width = \columnwidth]{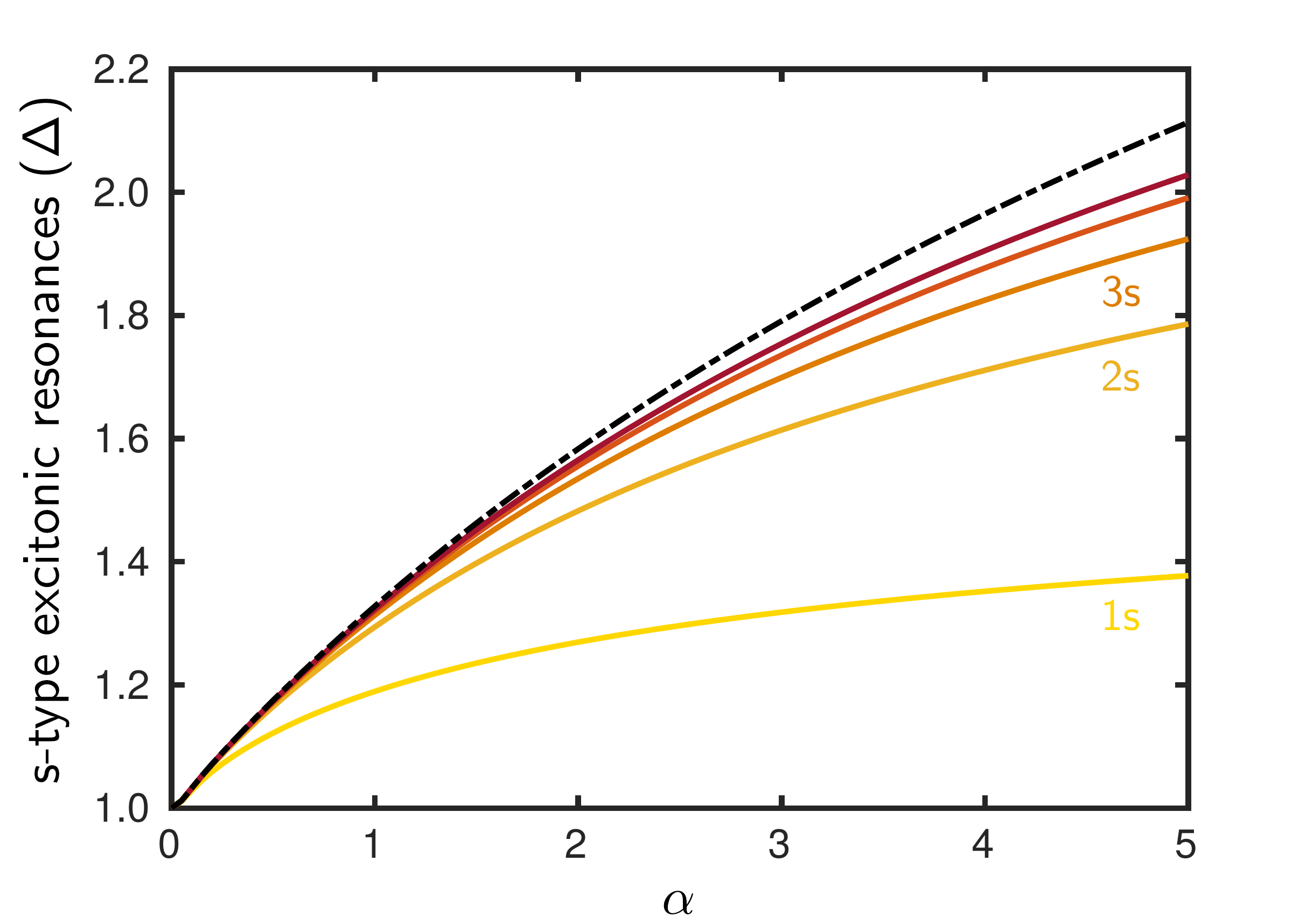} 
  \caption{\label{fig:alphaDependenceResonances} Lowest s-type energy eigenvalues of the Dirac-Wannier equation as a function of $\alpha$  
  with respect to the renormalized single-particle dispersion. The black dotted line indicates the renormalized band gap. The effective thickness parameter has been set to 
  $d=1.0\lambda_{s\tau}$. }
\end{figure}

The combined solution of the gap equations (\ref{Eq:Gap}) together with the DWE (\ref{WannierDirac}) allows us to determine the energetic positions of the excitonic resonances in an optical spectrum. In Fig. \ref{fig:alphaDependenceResonances}, we show the results for the five lowest $s$-type excitonic states for a fixed thickness $d=\lambda_C$ as function of coupling strength $\alpha$. 
For reference, we also plot the variation of the renormalized bandgap at one of the Dirac points (black dotted line). As expected,  the  binding energies increase with increasing Coulomb coupling strength. However, the increased binding is overcompensated by the bandgap renormalization, leading to an overall blue shift of the 
excitonic resonance spectrum. In the limit of strong Coulomb coupling, the increase of the $1s$-exciton binding energy is almost canceled by the renormalization of the bandgap, such that the lowest exciton resonance depends only weakly on the coupling strength.

\begin{figure}[hbt]
  \includegraphics[width = \columnwidth]{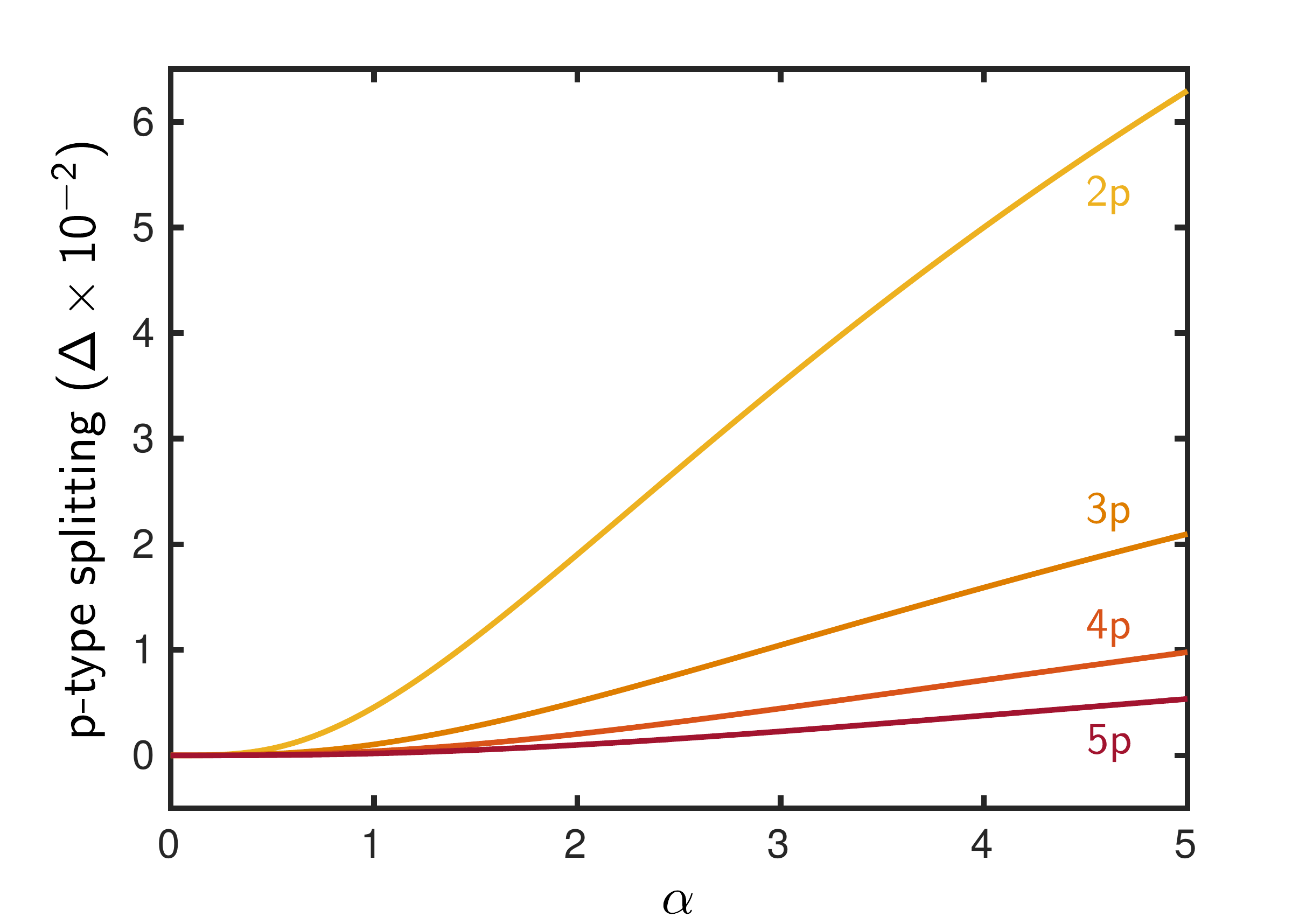}
  \caption{\label{fig:finestructure} Fine structure of the excitonic spectra, illustrated by the splitting of the lowest p-type excitonic states.}
\end{figure}

The coupling between $\phi$ and $\phi^*$ in the DWE leads to a fine structure in the exciton spectrum lifting the degeneracy between states with opposite orbital angular momentum.
In Fig. \ref{fig:finestructure}, we show the splitting of the lowest $p$-states for a fixed 
effective thickness $d=1.0\lambda_C$. In the limit of small values 
for the Coulomb coupling, the splitting increases quadratically switching over to a linear increase for large values of $\alpha$, respectively. For a suspended monolayer 
($\alpha\approx 3.0-5.0$), the splitting of the $2p$ states can be as high as $5-6\%$ of the noninteracting energy gap.  For supported monolayers, e.g. on a SiO$_2$ substrate 
($\alpha\approx 1.2-2.0$), our calculations predict a splitting on the order of $10-15$ meV, depending on the noninteracting gap of the specific material and on the screening. This value
 should be in the experimentally accessible range. 

\section{Multilayer Structures}\label{sec:MLtobulk}

So far, we investigated the excitonic scaling properties and the influence of finite layer thickness within a simplified model for the dielectric environment. In this section, we extend this approch and numerically study the properties of a multilayer TMDC system using the full solution of Poisson's equation within the anisotropic dielectric environment for the example of MoS$_2$.

For the MDF material parameters, we use the values given in Ref. \onlinecite{xiao2012},  $\Delta_A=1.585$ eV, $\Delta_B=1.735$ eV,  $\alpha^{[0]}=e^2/\hbar v_F=e^2/ta=4.11$,  from which we obtain the Compton wavelengths $\lambda_A=4.432$ \AA,\ $\lambda_B=4.049$ \AA,\  and the screening length $r_0=11.62$ \AA.\  To determine the Coulomb potential, we take the bulk in-plane and out-of-plane dielectric constants from Ref.\onlinecite{ghosh2013}, $\epspar^B=8.29$ and $\epsz=3.92$. Using a layer-to-layer-distance $D=6.2$ \AA,\ we find a background contribution to the in-plane dielectric constant $\epspar=4.54$.

\begin{figure}[hbt]
\includegraphics[width = \columnwidth]{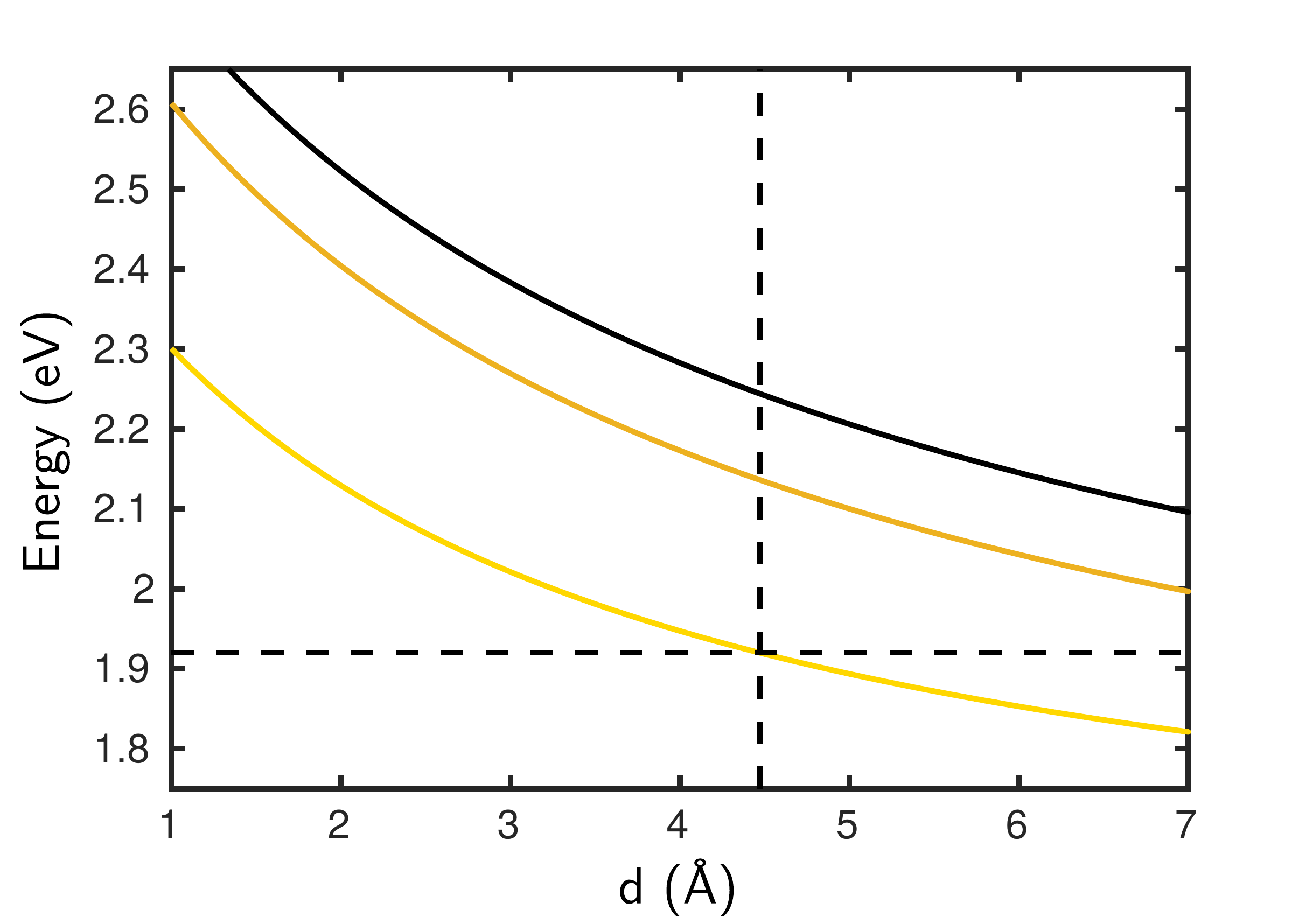}
  \caption{Predicted resonance positions for MoS$_2$ on SiO$_2$ as function of effective thickness. The solid lines show the theoretical position of the 1s (yellow), 2s (dark-yellow) resonance and the gap (black), the dashed lines mark the values $E=1.92$ eV and the best fit for the thickness parameter.\label{fig:MoS2onSiO2}}
\end{figure}

In a first step, we fix the only undetermined parameter in our theory, namely the effective thickness parameter $d$. To this end, we plot the renormalized gap and exciton resonances as funtion of $d$ and compare the resulting predictions with experimentally available data.
In Fig. \ref{fig:MoS2onSiO2}, we show the result of this procedure for the example of MoS$_2$ on SiO$_2$, where we use a constant dielectric constant $\epsilon_S=3.9$ for the SiO$_2$ substrate.

We fit the effective thickness parameter such that we obtain $E=1.92$ eV as the energy of the lowest exciton resonance, which is in the range of  measured values \cite{kioseoglou2012,mak2013,mitioglu2016}. As can be recognized, best agreement is obtained for an effective thickness parameter  $d=4.47$ \AA\ which is smaller than the layer separation $D$.  The corresponding values for the bandgap and the first excited exciton resonance are then $E_G=2.244$ eV and $E_{2s}=2.136 $ eV,  
 giving binding energies of $E^B_{1s}=324$ meV and $E_{2s}^B=108$  meV for MoS$_2$ on SiO$_2$ respectively.

Unfortunately, as the value for the gap is difficult to determine experimentally, we cannot directly compare  the findings for the bandgap and exciton binding energy with experiment. 
However, we can use the optimized value for the effective thickness to predict the bandgap and exciton resonances for a suspended monolayer, yielding $E_G=2.55$ eV and $E_{1s}=1.96$ eV, and a binding energy for the $1s$-exciton of  $E_{1s}^B=0.599$ eV. 
These values are in pretty good agreement with the values of $E_G=2.54$ eV and $E_{1s}^B=0.63$ eV reported in Ref.\onlinecite{qiu2016}.


\begin{figure}[hbt]
 {\centerline{
  {\includegraphics[width = \columnwidth]{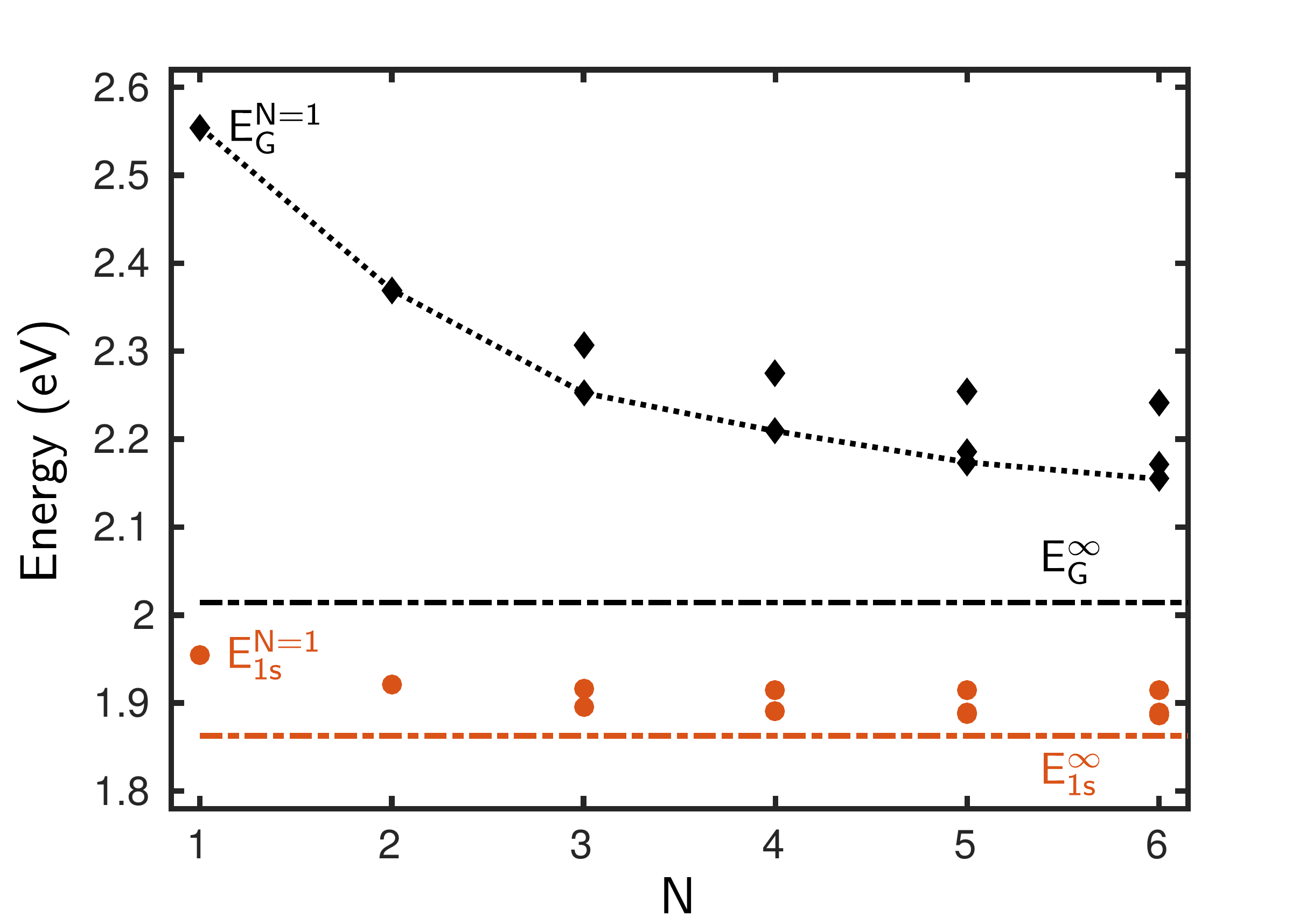}}
}}
  \caption{Left: 
Renormalized free-particle transition energies and excitonc positions as function of total layer number $N$ for suspended MoS$_2$ using $d=4.47 \AA\ $. With increasing layer number the number of bands increases accordingly and are classified by the layer index $n=1,\dots N$  (see text for explanation). The black diamonds show the transition energies $E_G(n,n)$ and the orange dots the resonance positions $E_{1s}(n,n)$ in the individual layers and the solid lines their weighted averages. 
The dashed lines indicate the bulk limit $N\rightarrow\infty$.
\label{fig:MoS2dependenceonN}}
\end{figure}

Once the thickness parameter is fixed, we are able to compute the renormalized bands and resonance positions for samples with arbitrary layer number and substrates. 
If we increase the number of layers, the number of bands within the first 2D Brillioun-zone is increased accordingly.  For the effective 2D quasi-particles that are localized well within a given layer, we can use the layer number $n$ within the stack as a good quantum number. In the following, we introduce the notation $E_G(n,m)=E^{c}_{n\bq=0}-E^{\nu}_{m\bq=0}$ for the transition energy between the top of the $n^{th}$ valence band and the bottom of the
 $m^{th}$ conduction band at the $K$-points, and a similar notation for the exciton resonances.

In Fig. \ref{fig:MoS2dependenceonN}, we show the variation of the renormalized valence-to-conduction band transition energies $E_G(n,n)$ and of the corresponding lowest exciton resonances $E_{1s}(n,n)$ with increasing number of layers. Since the effective local dielectric functions differ for different layers in the sample, both the transition energies and the excitonic resonances between bands  associated with different layers are non-degenerate,  leading to additional resonances in  the optical spectra of the multilayer structure.  For each value of $N$, the dots denote the transition energies $E_G(n,n)$ and $E_{1s}(n,n)$ for $n=1,N$, and the lines represent their weighted average. In the bulk limit $N\rightarrow\infty$, we find  $E_G^{\infty}=2.03$  and  $E_{1s}^{\infty}=1.88$, giving a binding energy of 150 meV for the lowest lying bulk exciton. These values are in good agreement with {\it GW}-BSE based {\it ab initio} results reported in Ref.\onlinecite{komsa2012}, where a binding energy of 130 meV was found for the bulk $A$-exciton. For reference, the  respective bulk limits for the band gap and lowests exciton are indicated in Fig.  \ref{fig:MoS2dependenceonN} by the dashed lines.


\begin{figure}[hbt]
 \includegraphics[width = \columnwidth]{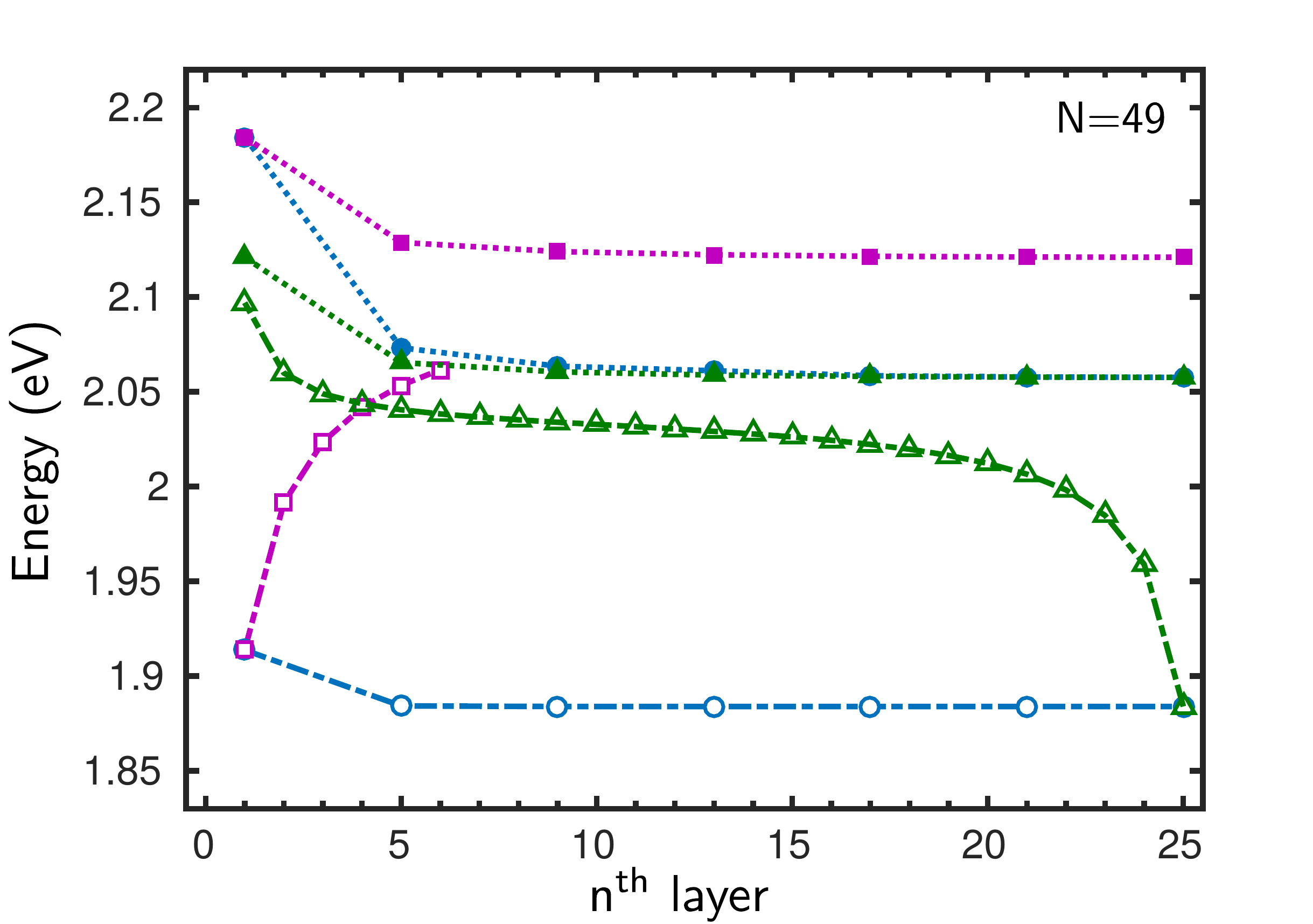}
  \caption{Transition energy $E_G(n,n)$  between the conduction band bottom and top of the valence band for an electron and hole localized in the same layer (green dots), an electron located in the top layer and a hole in the $n^{th}$ layer ($E_G(1,n)$, pink squares), an electron in the middle layer and hole in the $n^{th}$ layer ($E_G(n,25)$, green triangles) and similar for the repective intra- and interlayer excitons (open symbols). 
\label{fig:intraandinterlayer}}
\end{figure}

Besides their intralayer interaction, the electrons and holes in a multilayer structure interact also with carriers in neighboring layers with the possibility to form bound interlayer excitons. To illustrate these features, we plot in Fig. \ref{fig:intraandinterlayer} the free-particle transition energies $E_G(n,n)$ and  resonance energies $E_{1s}(n,n)$ for intralayer excitons where the electron-hole pair resides within the same layer, as well as the
interlayer transition energies $E_G(1,n)$  and $E_G(n,25)$, and energies of interlayer excitons $E_{1s}(1,n)$  and $E_{1s}(n,25)$ where an electron is confined in the $n^{th}$ layer and the hole in top or middle layer, respectively. We see that the interlayer excitons form a whole spectral series with decreasing binding energy for increasing spatial electron-hole separation. 

Due to our model assumption of electronically independent layers, the interlayer excitons are optically dark and cannot be observed in optical spectra. However, if we relax the assumption of electronically fully independent layers but allow for a finite overlap of the electron and hole wave functions in different layers, these interlayer excitons gain a finite oscillator strength. Assuming Gaussian distributions for the electron and hole densities, we can estimate the electron-hole overlap between different layers from the integral $\left|\int dz \phi_e(z)\phi_h(z-nD)\right|^2$ determining the oscillator strength for the respective interlayer excitons.

\begin{figure}[hbt]
 \includegraphics[width = \columnwidth]{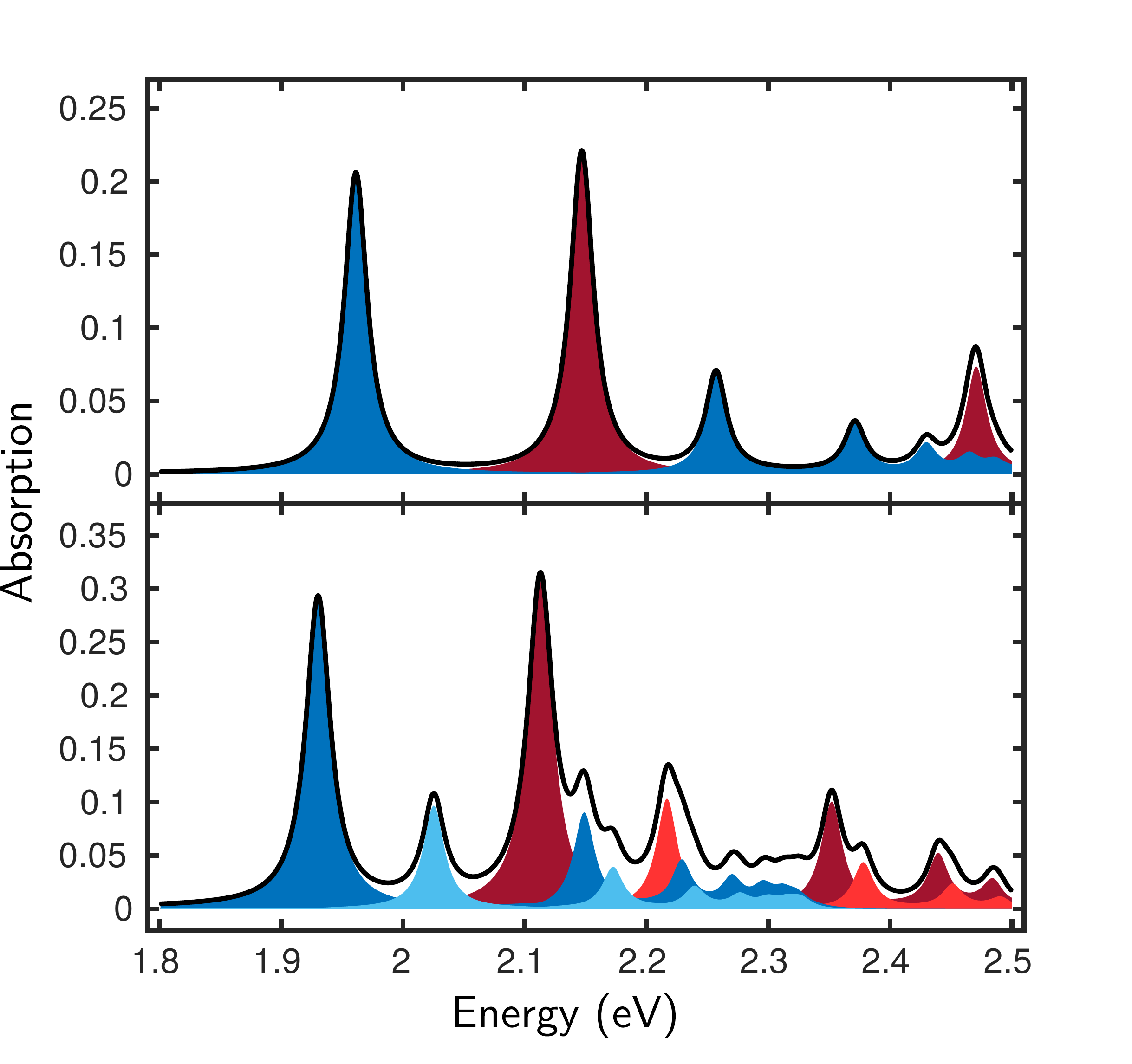}
  \caption{Calculated absorption spectra of a suspended mono- and bilayer MoS$_2$. The black lines show the total absorption spectra, whereas the dark blue and dark red fillings correspond to the $A$ and $B$ intralayer contributions, and the light blue and light red fillings to the $A$ and $B$  interlayer contributions. Spectra have been calculated using a nonradiative homogeneous linwidth $\hbar\gamma=10$ meV.
\label{fig:MoS2MLundBLspektren}}
\end{figure}

Using these model assumptions, we can compute optical absorption spectra for different multilayer systems. In Fig. \ref{fig:MoS2MLundBLspektren}, we show the results for a suspended mono- and bilayer MoS$_2$, using the screened Coulomb potential and thickness $d=4.47$ \AA\ .  The signature in the spectral range between the lowest $A$ and $B$ excitons, that are red shifted by roughly $30$ meV, is the lowest interlayer exciton. Furthermore, we see a clear red shift of the intralayer excitons in the bilayer relative to the monolayer. 

\begin{figure}[hbt]
 \includegraphics[width = \columnwidth]{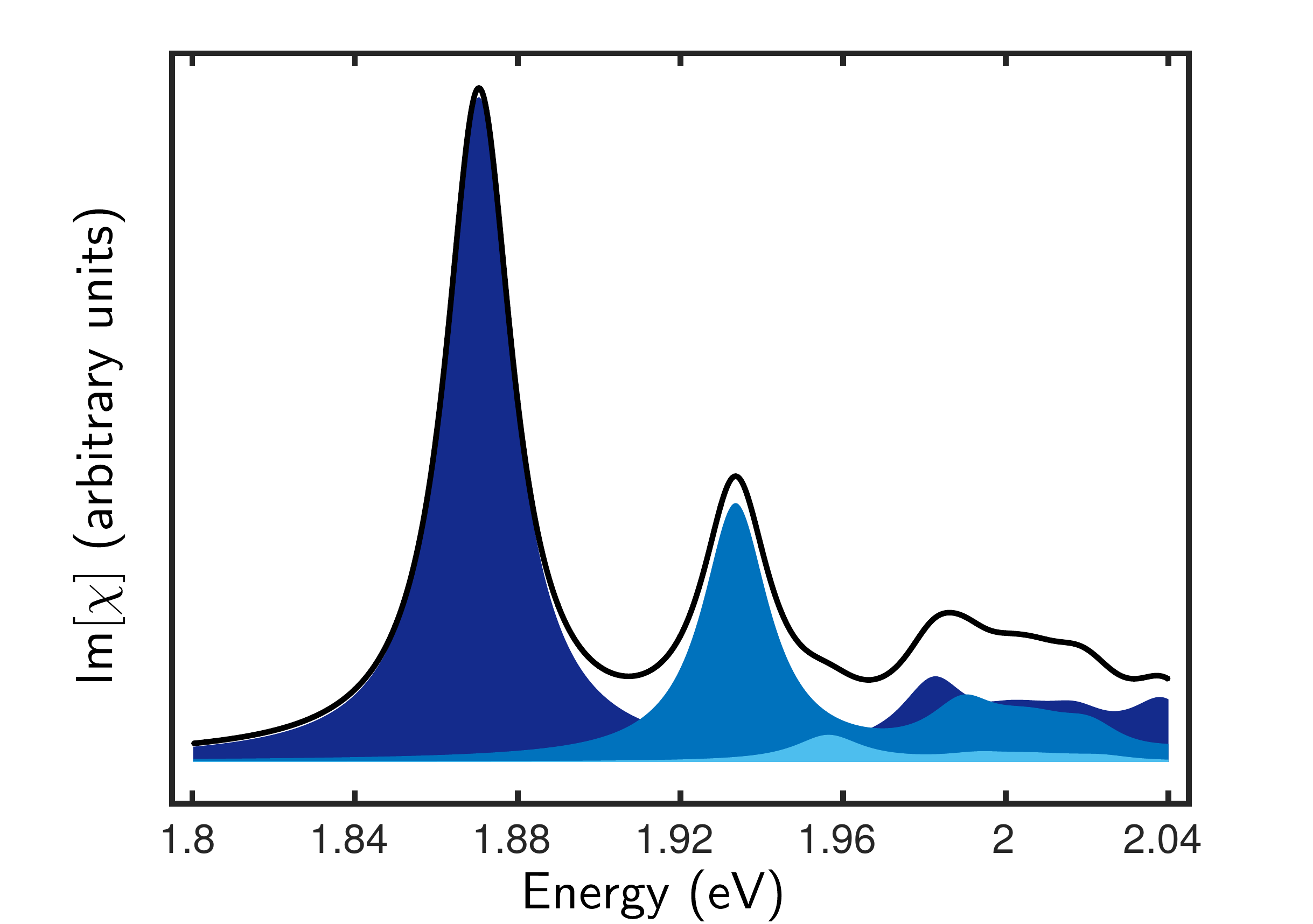}
  \caption{Imaginary part of the linear susceptibility of MoS$_2$ in the bulk limit using a  homogeneous linwidth $\hbar\gamma=10$ meV.  The black line shows the imaginary part of the total linear susceptibility, whereas the dark blue filling shows the intralayer contributions, the blue filling the next neighbor interlayer, and the light blue filling the next-next nearest neighbour intralayer excitons  contributions respectively.
\label{fig:MoS2bulk}}
\end{figure}

In Fig. \ref{fig:MoS2bulk}, we show the spectrum for a multilayer sample in the limit $N\rightarrow\infty$ in the spectral region of the $A$-exciton resonance series. The dominant peak at $E=1.87$ eV and the absorption features  slightly below the gap (at $2.03$ eV) correspond to the $A$-intralayer exciton series. The pronounced feature around $E=1.93$ eV results from the next-neighbor interlayer exciton, where electrons and holes are confined in neighboring layers. 
 
 It is interesting to compare these predictions with experimental findings on {\it bulk} MoS$_2$ for which the absorption spectrum has been measured already in the 1970s\cite{beal1972,bordas1973,fortin1975}. Transitions that were associated with the $A$-exciton at the $K$-points of the Brillioun zone
have been observed around $1.92$, $1.96$ and $1.99$ eV. In the original publication, the resonance features were interpreted as groundstate and excited state transitions of a single exciton series.
However, neither the resonance positions nor the oscillator strength agree with the expectations based on an anisotropic 3D Rydberg series. These deviations have been discussed in the literature and have been explained by so called "central-cell corrections". 

The remarkable agreement of the spectral signatures in Fig. \ref{fig:MoS2bulk} with the measured resonances suggests the reinterpretation of the bulk exciton series as 2D intra- and interlayer excitons, despite some small deviations in the absolute positions of the dominant absorption peaks.
This interpretation is further supported by recent measurements on bulk MoS$_2$\cite{saigal2016}, where a bias-dependent  relative oscillator strength between the two dominant features
has been observed, indicating a distinct $z$-dependence of both signatures.

\section{Discussion}

In conclusion, we present a theoretical framework that allows us to compute the bandgap renormalization and $K$-point excitonic resonances of TMDC mono- and multilayer structures. Our method contains the effective monolayer thickness as undetermined parameter.  For the example of MoS$_2$, we show that by fitting this single parameter to obtain agreement for the lowest exciton resonance of a supported monolayer, we are able  to compute the bandgap and excitonic spectra of  samples with arbitrary layer numbers and substrates. In particular, we are able to predict the evolution of the bandgap and near-bandgap excitonic spectra over the whole range from monolayer to bulk. Our predictions for the bulk limit are in excellent agreement with experimental observations, suggesting a reinterpretation of the bulk $A$ and $B$ excitonic series in terms of effectively 2D intra- and interlayer excitons.

It is interesting, to compare our method with the  well established {\it GW}-BSE approach. In the  {\it GW}-BSE approach, the quasi-particle bandgap is computed from many-body perturbation theory on top of the DFT band structure. Subsequently, excitonic states are obtained as solution of the Bethe-Salpeter-equation (BSE). The major strength of the {\it GW}-BSE approach is that it is fully {\it ab initio}, and as such, free of any undetermined parameters. However, this comes at the price of being numerically very demanding.  The treatment of quasi-2D structures within  {\it GW}-BSE is computationally even more challenging, as it  requires large supercells to avoid spurious interactions between adjacent layers. 
The numerical complexity of the {\it GW}-BSE approach has not only lead to a wide range of reported  predictions for the bandgap and exciton bindings, it also limits its practical application to the description of groundstate and linear optical properties.
 
Methodically, our approach displays several similarities  to the {\it GW}-BSE approach. Similarly as {\it GW},  the gap equations provide a correction to the DFT bandstructure, and a subsequent solution of the Dirac-Wannier-equation within the renormalized bands gives access to the excitonic states. However, whereas the  {\it GW}-BSE equations involve many bands,
our approach is explicitly based on a two-band Hamiltonian, thus reducing the numerical cost enormously. Though an effective two-band Hamiltonian restricts the applicability of our method  to the simulation of the  near bandgap optical properties,   our method is extremely flexible to model different dielectric environments and  can be easily extended to describe nonlinear optical experiments.

Both qualitatively and quantitatively, our predictions are in very good agreement with well-converged {\it GW}-BSE based results\cite{qiu2016}. This, in addition to the excellent agreement  with experimental observations can be taken as strong indications that our model system captures the essential physics around the $K$-points of the Brillioun zone. In particular, we identify finite size effects as essentially responsible for the observed non-hydrogenicity not only of monolayer spectra, but also of multilayer spectra in the bulk limit. 

\begin{acknowledgements}
This work is a project of the Collaborative Research Center SFB 1083 funded by the Deutsche Forschungsgemeinschaft. We thank M. Rohlfing for stimulating discussions and for sharing his results on interlayer excitons in TMDCs prior to publication.
\end{acknowledgements}
\appendix

\begin{widetext}
\section{Solution of Poisson's Equation}\label{App:Poisson}

 \subsection{Bare Coulomb Interaction}

The 'bare'  Coulomb interaction corresponds to the Green function of Poisson's equation, i.e.,  is obtained as the solution of Eq. \ref{2DCoulomb} for the scalar potential 
with $\delta$-inhomogeneity $\rho(\bq_\parallel,z)=\delta(z-z')$ in the absence of a resonant polarization, but in the presence of the inhomogeneous, anisotropic background.
For a slab geometry consisting  of thickness $L=ND$ on a substrate with dielectric constant $\epsilon_S$,
we  have a spatial profile of the background dielectric tensor: 
\begin{eqnarray*}
\epspar(z)&=&\left\{\begin{array}{ll}1&z<0,\\
\epspar& 0<z<L,\\
\epsilon_S&L<z\end{array}\right.
\quad
\epsz(z)=\left\{\begin{array}{ll}1&z<0,\\
\epsz& 0<z<L,\\
\epsilon_S&L<z.\end{array}\right.
\end{eqnarray*}

Within the slab, the resulting  Coulomb potential of a point charge  located at $0<z'<L$ is given by
\begin{eqnarray}
V_\bq(z,z')&=&\frac{2\pi }{\kappa q}\left({\rm e}^{-\sqrt{\frac{\epspar}{\epsz}}q_\parallel |z-z'|}
+
c_1{\rm e}^{-\sqrt{\frac{\epspar}{\epsz}}q_\parallel (z+z')}\right.
+c_2{\rm e}^{-\sqrt{\frac{\epspar}{\epsz}}q_\parallel (2L-z-z')}
+c_3
{\rm e}^{-\sqrt{\frac{\epspar}{\epsz}}q_\parallel (2L-z+z')}
+\left. c_3{\rm e}^{-\sqrt{\frac{\epspar}{\epsz}}q_\parallel (2L+z-z')}
\right)\nonumber\\
\label{Eq:V0multilayer}
\end{eqnarray}
with 
\end{widetext}

\begin{eqnarray*}
\kappa&=&\sqrt{\epspar\epsz},\\
c_1&=&\frac{(\kappa+\epsilon_S)(\kappa-1)}{{\cal{N}}
},\\
c_2&=&\frac{(\kappa-\epsilon_S)(\kappa+1)}{{\cal{N}}},
\\
c_3&=&\frac{(\kappa-\epsilon_S)(\kappa-1)}{{\cal{N}}
},\\
{\cal{N}}&=&\left(\kappa+\epsilon_S\right)\left(\kappa+1\right)-\left(\kappa-\epsilon_S\right)\left(\kappa-1\right){\rm e}^{-2\sqrt{\frac{\epspar}{\epsz}}q_\parallel L}.
\end{eqnarray*}
\begin{widetext}
In Eq. (\ref{Eq:V0multilayer}), the first term describes the direct interaction between the two point charges, the second term interaction of the point charge at $z$ with the image charge of $z'$ from the vacuum/multilayer interface, the third term correspondingly from the multilayer/substrate interface and the last term the interaction between image charges from both interfaces.
Interaction with higher order image charges are contained in the denominator ${\cal {N}}$.

Relevant for the intralayer exciton and band gap renormalization is the intralayer Coulomb potential  $V_q(z_n,z_n)$ with $z_n=(n-1/2)D$:
\[
V_\bq(z_n,z_n)=\frac{2\pi }{\kappa q}\left(1
+
c_1{\rm e}^{-\sqrt{\frac{\epspar}{\epsz}}q_\parallel (2n-1)D}+c_2{\rm e}^{-\sqrt{\frac{\epspar}{\epsz}}q_\parallel 2(N-n-1/2)D}+2 c_3
{\rm e}^{-\sqrt{\frac{\epspar}{\epsz}}q_\parallel 2L}
\right)
\]

For $\sqrt{\frac{\epspar}{\epsz}}q_\parallel L\ll 1$, the intralayer Coulomb potential reduces to $V=4\pi/(\epsilon_S+1)q_\parallel$, i.e., to the vacuum 2D Coulomb interaction screened by substrate screening only, while if  $\sqrt{\frac{\epspar}{\epsz}}q_\parallel L\gg1$,
it reduces to 
\[
\frac{2\pi}{\kappa q_ \parallel}\left(1+\frac{\kappa-1}{\kappa+1}{\rm e}^{-\sqrt{\frac{\epspar}{\epsz}}q_\parallel 2(n-1/2)D}+
\frac{\kappa-\epsilon_S}{\kappa+\epsilon_S}{\rm e}^{-\sqrt{\frac{\epspar}{\epsz}}q_\parallel (2(N-n-1/2)D}
\right).
\]

In the left part of  Fig. \ref{epsilonfull}, we show the local dielectric functions for the middle layer of a suspended MoS$_2$ sample consisting of 1, 3, and 49 layers. At small wavenumbers, the 
 dielectric function of the middle layer can be approximated by a first order Taylor expansion, giving
\[
\epsilon(q)\approx \frac{\epsilon_S+1}{2}+N\frac{2\epspar\epsz-\epsilon_S^2-1}{4\epsz}qD.
\] 
The linear approximations corresponds to a Keldysh potential\cite{keldysh1979,cudazzo2011,berkelbach2013} with background screening $(\epsilon_S+1)/2$ and screening length $r=N\frac{2\epspar\epsz-\epsilon_S^2-1}{2\epsz(\epsilon_S+1)}D$. However, the linear approximation breaks down if $qND>1$, where the dielectric function approaches its bulk value. Estimating the relevant $q$-values by the inverse exciton radius $r_X$ (note: the exciton radius should not be interchanged with the exciton Bohr radius; only for hydrogen-like excitons these values coincide), this means that the total sample dimensions should not exceed the in-plane exciton radius. While this condition may hold for a monolayer, it is clearly invalid for a multilayer structure with large layer numbers. As can be recognized in  Fig. \ref{epsilonfull}, in a sample with 49 layers the nonresonant dielectric function jumps to its bulk background value at infinitesimal $q$-values.

 \subsection{Screening}

\begin{figure}
\includegraphics[width=.3\textwidth]{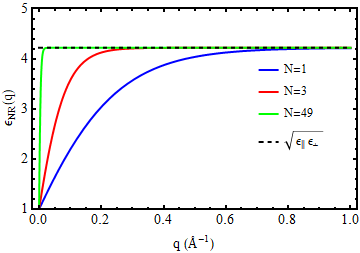}
\includegraphics[width=.3\textwidth]{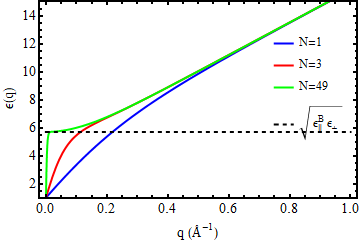}
\includegraphics[width=.3\textwidth]{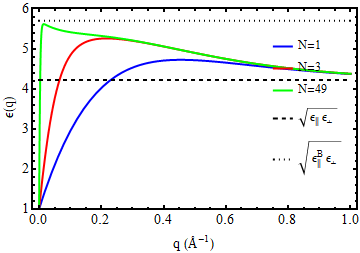}
\caption{Left: Nonresonant contributions to the dielectric  function for the middle layer of a suspended MoS$_2$ sample consisting of $N$ layers. Middle: Total effective dielectric function for middle layer of a suspended MoS$_2$ sample with $N$ layers within the strict 2D limit. Right :Total effective dielectric function for middle layer of a suspended MoS$_2$ sample with $N$ layers including finte size effects with an effective layer thickness $d=4.3$ \AA. 
\label{epsilonfull}}
\end{figure}

Within linear response theory, the  polarization  in an inhomogeneous medium   induced by an external perturbation field $\phi$ can be expressed in terms of a nonlocal susceptibility
\[
\P_L(\bq,z,\omega)=-ie^2\bq\int dz'\chi_L(\bq,z,z',\omega)\phi(\bq,z',\omega)
\]
where the $z$-dependence of the susceptibility reflects the spatial profile of the induced carrier density. For the multilayer system,  we assume charge distributions well localized within the layers, such that the integration region can be restricted to a region of thickness $D$ around the layer centers:
\[
\P_L(\bq,z,\omega)=-ie^2\bq\sum_{n=1}^N\rho_\bq(z-z_n)\chi_L(\bq,\omega)\bar\phi^n(\bq,\omega)
\]
where
\[
\bar\phi^n(\bq,\omega)=\int_{-D/2}^{D/2} dz'\rho_{-\bq}(z'-z_n)\phi(\bq,z',\omega).
\]
In the strict 2D limit, this corresponds to Ansatz \ref{Eq:PolAnsatz} of the main text.
The formal solution of equation \ref{2DCoulomb} with charge distribution $\rho_{ext}(\bq,z)$ is than given by
\begin{eqnarray}
\phi(\bq,z,\omega)&=&\phi_{ext}(\bq,z,\omega)
-e^2q^2\sum_n\chi_L(\bq,\omega)\int d z'V_\bq(z,z')\rho_\bq(z'-z_n)\bar\phi^n(\bq,\omega)\nonumber\\
&\approx&\phi_{ext}(\bq,z,\omega)
-e^2q^2\sum_n\chi_L(\bq,\omega)\int_{-D/2}^{D/2} d z'V_\bq(z,z')\rho_\bq(z'-z_n)\bar\phi^n(\bq,\omega)
\label{phiformal}
\end{eqnarray}
where $V_\bq(z,z')$ is the Coulomb interaction screened by the anisotropic background given in Eq. \ref{Eq:V0multilayer} and
\[
\phi_{ext}(\bq,z,\omega)=\int dz'V_\bq(z,z')\rho_{ext}(\bq,z')
\]
is the potential of the external charge distribution. Multiplication of Eq. (\ref{phiformal}) with $\rho_{-\bq}(z-z_m)$ and integration over $z$ gives
\begin{eqnarray}
\bar\phi^m(\bq,\omega)&=&\bar\phi_{ext}^m(\bq,\omega)
-e^2q^2\sum_n\chi_L(\bq,\omega)\bar V_\bq^{mn}\bar\phi^n(\bq,\omega)
\label{phiformalquasi2D}
\end{eqnarray}
with the quasi-2D bare Coulomb potential
\[
\bar V_\bq^{mn}=\int_{-D/2}^{D/2}dz \int_{-D/2}^{D/2}dz'\rho_{-\bq}(z-z_m)V_\bq(z,z')\rho_\bq(z'-z_n).
\]
The solution of  Eq. \ref{phiformal} can be obtained by a matrix inversion:
\begin{eqnarray}
 \bar\phi^m(\bq,\omega)=\sum_l\left(\delta_{ml}+e^2q^2\chi_L(\bq,\omega)\bar V_\bq^{ml}\right)
 ^{-1}\bar\phi_{ext}^l(\bq,\omega),
\label{phimatrix}
\end{eqnarray}
and the screened Coulomb interaction given in Eq. (\ref{VCmatrix}) in the main text is obtained by choosing $\rho_{ext}(\bq,z)=\delta(z-z_n)$.

For a monolayer in the strict 2D limit, the solution simplifies to
\begin{eqnarray}
\phi^{2D}(\bq)&=&\frac{\phi_{ext}(\bq,z=D/2)}{1+ e^2q^2 
\chi_L(\bq,\omega)V_\bq(D/2,D/2)}
\end{eqnarray}
where $\phi^{2D}$ is the screened external potential.  This result generally depends on the slab 
thickness $D$ and becomes inedpendent of $D$ only in the two limiting cases $D\rightarrow 0$ and $D\rightarrow\infty$. The limit $D\rightarrow 0$ correponds to a monolayer on a substrate, whereas the limit $D\rightarrow\infty$ corresponds to a monolayer embedded in a homogeneous anisotropic medium. Defining $\epsilon_{\rm eff}$ by $\epsilon_{\rm eff}=(\epsilon_S+1)/2$ and $\epsilon_{\rm eff}=\sqrt{\epsz\epspar}$ respectively,
the localized 2D polarization contributes to the longitudinal dielectric function according to
$\epsilon_{\rm RES}=1+2\pi e^2 q_\parallel\chi_L(\bq,\omega)/\epsilon_{\rm eff}$. 
If the 2D susceptibility is independent of $\bq$ and $\omega$, this part again corresponds to the Keldysh potential, with a resonant contribution to the anti-screening length 
$r_0=2\pi e^2\chi_L/\epsilon_{\rm eff}$.

In the middle part of  Fig. \ref{epsilonfull}, we show the resulting total effective dielectric function of the middle layer of a suspended multilayer sample, where we treat the resonant contribtions to the dielectric function in the long wavelength limit. As can be recognized,  if $N$ is increased, the longwavelength limit of the total dielectric function $\epsilon(q=0)$ approaches the bulk value $\sqrt{\epspar^B\epsz}$, with an in-plane component corresponding to the (fully screened) DFT bulk value, whereas the  monolayer dielectric function in the small $q$ regime can again be approximated by a Keldysh potential with a total linear coefficient $r_{\rm tot}=r+2r_0/(\epsilon_S+1)$. However, whereas the nonresonant contribution does not exceed the bulk back-ground value $\sqrt{\epspar\epsz}$, the total dielectric function increases linearly, exceeding the DFT fully screened bulk value by far. This unphysical result results from the strict 2D treatment of the carriers, and the invalidity of the long-wave-length limit for the polarization function in this regime.
In the right  part of  Fig. \ref{epsilonfull}, we show the total effective dielectric function including finite size effects by the Ohno potential. As can be recognized, the effective dielectric function for the middle layer  increases linearly for small $q$-values starting at $\epsilon(q=0)=1$. However, due to finite size effects, the dielectric function does not exceed the fully screened bulk limit $\sqrt{\epspar^B\epsz}$, but reaches a maximum value between the bulk back-ground value  $\sqrt{\epspar\epsz}$ and  the fully screened DFT bulk limit $\sqrt{\epspar^B\epsz}$. The $q$-value at wich the maximum is achieved decreases with increasing number of layers, nicely reproducing the bulk long wave-length limit for large layer numbers. 

Finally, we compare the effective dielectric function for the monolayer with two recent publications whrere  the effective 2D dielectric function for a monolayer TMDC has been exctracted from a first principles supercell calculation, once using a dielectric model similar to ours\cite{latini2015}, that accounts for finite size effects, and once using a truncated Coulomb potential\cite{qiu2016}. Both approaches find a dielectric function starting at $\epsilon(q=0)=1$, and a maximum value in the region $q\approx 0.3 \AA^{-1}$. At large $q$-values, the dielectric function decreases to $\epsilon(q\rightarrow\infty)=1$ again, reflecting the lack of dielectric screening at small distances. Apparently, our modell overestimates the effect of screening in the large $q\gg 1/d$  limit. This is a consequence of using a {\it constant} background dielectric constant, which is inappropriate for large $q$-values. Indeed, choosing a background dielectric constant $\epspar=\epsz=1$ in our model and lumping the back-ground contributions into a linear coefficient $r_{\rm tot}=r+r_0$ instead, the monolayer dielectric function is in good agreement with both Refs. \onlinecite{latini2015,qiu2016}.
 On the other hand,
 our model system is in good agreement with the findings in Refs. \cite{latini2015,qiu2016} in the region $q\lesssim 1/d$, relevant for excitons in the Wannier-limit, and
 produces the correct bulk limit if the number of layers is increased. In contrast,  lumping the back-ground contributions into the linear increase in the small $q$-region produces a wrong  bulk limit  $\kappa^L(q\rightarrow 0)=\sqrt{1+2r/L}$, which can be applied to bulk ($L=D$) as well as to a supercell calculation with supercell period $L$. 
\end{widetext}



%

\end{document}